\documentclass[preprint]{elsarticle}

\usepackage{lineno,hyperref}
\modulolinenumbers[5]

\usepackage{graphicx}
\usepackage{amsmath,amsfonts,amssymb,amsthm}
\usepackage{float,dsfont,url,color}
\usepackage{multirow}
\usepackage{ulem}
\usepackage{booktabs}
\newcommand{\bs}{\boldsymbol}

\begin{document}
\begin{frontmatter}

\title{Overfitting Bayesian Mixtures of Factor Analyzers with an Unknown Number of Components}
\author{Panagiotis Papastamoulis}
\address{University of Manchester\\Faculty of Biology, Medicine and Health\\Division of Informatics, Imaging and Data Sciences\\Michael Smith building, Oxford Road, M13 9PL, Manchester, UK}
\ead[url]{panagiotis.papastamoulis@manchester.ac.uk}

\begin{abstract}
Recent advances on overfitting Bayesian mixture models provide a solid and straightforward approach for inferring the underlying number of clusters and model parameters in heterogeneous datasets. The applicability of such a framework in clustering correlated high dimensional data is demonstrated. For this purpose an overfitting mixture of factor analyzers is introduced, assuming that the number of factors is fixed. A Markov chain Monte Carlo (MCMC) sampler combined with a prior parallel tempering scheme is used to estimate the posterior distribution of model parameters. The optimal number of factors is estimated using information criteria. Identifiability issues related to the label switching problem are dealt by post-processing the simulated MCMC sample by relabelling algorithms. The method is benchmarked against state-of-the-art software for maximum likelihood estimation of mixtures of factor analyzers using an extensive simulation study. Finally, the applicability of the method is illustrated in publicly available data. 
\end{abstract}


\end{frontmatter}


\section{Introduction}

Factor Analysis (FA) is a popular statistical model that aims to explain correlations in a high-dimensional space by dimension reduction. This is typically achieved by expressing the observed multivariate data as a linear combination of a smaller set of hypothetical and uncorrelated variables known as factors. The factors are not observed, so they are treated as missing data. The reader is referred to \cite{kim1978factor, bartholomew2011latent} for an overview of factor analysis models, estimation techniques and applications. 

However, when the observed data is not homogeneous, the typical FA model will not adequately fit the data. In such a case, a Mixture of Factor Analyzers (MFA) can be used in order to take into account the underlying heterogeneity. Thus, MFA models jointly treat two inferential tasks: model-based density estimation for high dimensional data as well as dimensionality reduction. Estimation of MFA models is straightforward by using the Expectation-Maximization (EM) algorithm \cite{Dempster:77, ghahramani1996algorithm, McLachlan:00, mclachlan2003modelling, mclachlan2011mixtures}. The family of parsimonious Gaussian mixture models (PGMM) is introduced in \cite{McNicholas2008,  mcnicholas2010serial, doi:10.1093/bioinformatics/btq498},  which is based on Gaussian mixture models with parsimonious factor analysis like covariance structures.  Under a Bayesian setup, \cite{Fokoue2003} estimate the number of mixture components and factors by simulating a continuous-time stochastic birth-death point process using a Birth-Death MCMC algorithm \citep{stephens2000}. Their algorithm is shown to perform well in small to moderately scaled multivariate data. 

Fully Bayesian approaches to estimate the number of components in a mixture model include the Reversible jump MCMC (RJMCMC) \citep{Green:95, Richardson:97,dellaportas2006multivariate, papRJ}, Birth-death MCMC (BDMCMC) \citep{stephens2000} and allocation sampling \citep{Nobile2007} algorithms. In recent years there is a growing progress on the usage of overfitted mixture models in Bayesian analysis \citep{rousseau2011asymptotic, overfitting}. An overfitting mixture model consists of a number of components  which is much larger than its true (and unknown) value. Under a frequentist approach, overfitting mixture models is not a recommended practice. In this case, the true parameter lies on the boundary of the parameter space and identifiability of the model is violated due to the fact that some of the component weights can be equal to zero or some components may have equal parameters. Consequently, standard asymptotic Maximum Likelihood theory does not apply in this case \citep{li1988mixtures}. Choosing informative prior distributions that bound the posterior away from unidentifiability sets can increase the stability of the MCMC sampler, however these informative priors tend to force too many distinct components and the possibility of reducing the overfitting mixture to the true model is lost (see Section 4.2.2 in \cite{fruhwirth2006finite}). Under suitable prior assumptions introduced by \cite{rousseau2011asymptotic}, it has been shown that asymptotically the redundant components will have zero posterior weight and force the posterior distribution to put all its mass in the sparsest way to approximate the true density. Therefore, the inference on the number of mixture components can be based on the posterior distribution of the ``alive'' components of the overfitted model, that is, the components which contain at least one allocated observation.

The simplicity of this approach is in stark contrast with the fully Bayesian approach of treating the number of clusters as a random variable. For example, in the RJMCMC algorithm the researcher has to design sophisticated move types that bridge models with different number of clusters. On the other hand, the allocation sampler is only applicable to cases where the model parameters can be analytically integrated out. Even in such cases though, the design of proper Metropolis-Hastings moves on the space of latent allocation variables of the mixture model is required to obtain a reasonable mixing of the simulated MCMC chain (see \cite{Nobile2007, papastamoulis2016bayesbinmix}). 

The contribution of this study is to utilise recent advances on overfitting mixture models \citep{overfitting} to the context of Bayesian MFA \citep{Fokoue2003}. We use a Gibbs sampler \citep{geman, gelfand} which is embedded in a prior parallel tempering scheme in order to improve the mixing of the algorithm. In addition, we explore the usage of information criteria for estimating the number of factors. After estimating the number of clusters and factors, we perform inference on the chosen model by dealing with identifiability issues related to the label switching problem \citep{papastamoulis2016label}. Our results indicate that overfitting Bayesian MFA models  provide a simple and efficient approach to estimate the number of clusters in correlated high-dimensional data.

The rest of the paper is organized as follows. Section \ref{sec:FA} reviews the basics of FA models. Finite mixtures of FA models are presented in Section \ref{sec:MFA} and a brief review of previous frequentist approaches is given in Section \ref{sec:previous}. The Bayesian formulation is presented in Section \ref{sec:priors}. The overfitting MFA model is introduced in Section \ref{sec:overfitting}. Section \ref{sec:qSelection} deals with estimating the number of factors using information criteria. Section \ref{sec:tempering} presents the prior parallel tempering scheme which is incorporated into the MCMC sampler. Identifiability issues related to the label switching phenomenon are discussed in Section \ref{sec:label} and further details of the overall implementation are given in Section \ref{sec:details}. Our method is illustrated and compared against the EM algorithm in Section \ref{sec:results} using a simulation study (Section \ref{sec:simulationStudy}) as well as three publicly available datasets (Sections \ref{sec:realData}, \ref{sec:wines} and \ref{sec:coffee}). The paper concludes in Section \ref{sec:discussion}. Further technical details and simulation results are provided in the Appendix.

\section{Methodology}\label{sec:method}

At first we introduce some conventional guidelines that will be followed in our notation throughout this paper, unless explicitly stated otherwise. We will use bold face for vectors and matrices. The notation $\alpha_k$ will correspond to the $k$-th member of a vector $\bs a$. In addition, $\bs{\mathrm{A}}_k$ will denote the $k$-th member of a vector $\bs{\mathrm{A}}$ whose elements are matrices. The $(i,j)$ element of a matrix $\bs \Sigma$ will be denoted by the corresponding lower case letter, that is, $\sigma_{ij}$. The transpose matrix of $\bs \Sigma$ will be denoted as $\bs \Sigma^T$. We will not differentiate the notation between random variables and their specific realizations. We use $f(x|y)$ to denote the probability mass or density function of $x$ given $y$. For a discrete random variable $z$, the notation $\mbox{P}(z = k)$ will be also used to denote the probability of the event $\{z = k\}$. The $p\times p$ identity matrix is denoted as $\bs{\mathrm{I}}_p$, $p \in \mathbb N$.


\subsection{Factor Analysis Model}\label{sec:FA}

Let $\bs x = (\bs x_1,\ldots,\bs x_n)$ denote a random sample of $p$ dimensional observations with $\bs x_i\in\mathbb R^{p}$; $i= 1,\ldots,n$. We assume that $\bs x_i$ is expressed as a linear combination of a latent vector (factors) $\bs x_i\in\mathbb R^{q}$
\begin{equation}\label{eq:fa}
\bs x_i = \bs\mu + \bs\Lambda \bs y_i + \bs\varepsilon_i. 
\end{equation}
The unobserved random vector $\bs y_i$ lies on a lower dimensional space, that is, $q < p$ and it consists of uncorrelated features $y_{i1}, \ldots,y_{iq}$. In particular, we assume that 
\begin{equation}\label{eq:y}
\bs y_i \sim \mathcal N_q(\bs 0,\bs{\mathrm{I}}_q),
\end{equation}
independent for $i = 1,\ldots,n$ and $\bs 0$ denotes a vector of zeros. The $p\times q$ dimensional matrix $\bs\Lambda = (\lambda_{rj})$ contains the factor loadings, while the $p$-dimensional vector $\bs\mu = (\mu_1,\ldots,\mu_p)$ contains the marginal mean of $\bs x_i$. For the error term $\bs\varepsilon_i$ assume that 
\begin{equation}\label{eq:same_sigma}
\bs\varepsilon_i \sim \mathcal N_p(\bs 0,\bs\Sigma)
\end{equation}
independent for $i = 1,\ldots,n$, where $\bs\Sigma = \mbox{diag}(\sigma_1^2, \ldots,\sigma_p^2)$. Furthermore, $\bs y_i$ is assumed independent of $\bs\varepsilon_i$; $i =1,\ldots, n$. It follows that
\begin{equation}\label{eq:x_given_y}
\bs x_i|\bs y_i \sim \mathcal N_p(\bs\mu+\bs\Lambda \bs y_i,\bs\Sigma).
\end{equation}
It can be easily derived that the marginal distribution of $\bs x_i$ is 
\begin{equation}\label{eq:x_marginal}
\bs x_i \sim \mathcal N_p(\bs\mu,\bs\Lambda\bs\Lambda^T + \bs\Sigma),
\end{equation}
independent for $i = 1,\ldots,n$.

According to the last expression, the covariance matrix of $\bs x_i$ is equal to $\bs\Lambda\bs\Lambda^T + \bs\Sigma$. As shown in Equation \eqref{eq:x_given_y}, the knowledge of the missing data ($\bs y_i$) implies that the conditional distribution of $\bs x_i$ has a diagonal covariance matrix. This is the crucial characteristic of factor analysis models, where they aim to explain high-dimensional dependencies using a set of lower-dimensional uncorrelated factors.

Given that there are $q$ factors, the number of free parameters in the covariance matrix $\bs\Lambda\bs\Lambda^T + \bs\Sigma$ is equal to $p + pq - \frac{1}{2}q(q-1)$ \citep{lawley1962factor}. The number of free parameters in the uncostrained covariance matrix of $\bs x_i$ is equal to $\frac{1}{2}p(p+1)$. Hence, under the factor analysis model, the number of parameters in the covariance matrix is reduced by
$$\frac{1}{2}p(p+1) - \left[p + pq - \frac{1}{2}q(q-1)\right] = \frac{1}{2}\left[(p-q)^2-(p+q)\right].$$
The last expression is positive if $q < \phi(p)$ where $\phi(p):=\frac{2p + 1 - \sqrt{8p+1}}{2}$, a quantity which is known as the Ledermann bound \citep{ledermann1937rank}. We assume that the number of latent factors does not exceed $\phi(p)$. When $q < \phi(p)$ it can be shown that $\bs\Sigma$ is almost surely unique \citep{BEKKER1997255}. Note however that this does not necessarily imply that the model is not identified if $q>\phi(p)$ \citep{bekker2014identification}. 

Given identifiability of $\bs\Sigma$, a second source of identifiability problems is related to  orthogonal transformations of the matrix of factor loadings. Indeed, consider a $p\times q$ matrix $\bs\Gamma$  with $\bs\Gamma^T\bs\Gamma = \bs{\mathrm{I}}_q$ and $\bs\Gamma\bs\Gamma^T = \bs{\mathrm{I}}_p$ (orthogonal matrix) and define $\tilde{\bs y}_i = \bs\Gamma \bs y_i$. It follows that the representation $\bs x_i = \bs\mu+\bs\Lambda\bs\Gamma^{T}  \tilde{\bs y_i} + \bs\epsilon_i$ leads to the same marginal distribution of $\bs x_i$ as the one in Equation \eqref{eq:x_marginal}. Following \cite{Fokoue2003}, we preassign values to some entries of $\bs\Lambda$, in particular we set the entries of the upper diagonal of the first  $(q-1)\times(q-1)$ block matrix of $\bs\Lambda$ equal to zero. 

Another identifiability problem is related to the so-called ``sign switching'' phenomenon, see e.g.~\cite{conti2014bayesian}. Observe that Equation \eqref{eq:fa} remains invariant when simultaneously switching the signs of a given row $r$ of $\bs\Lambda$; $r = 1,\ldots,p$ and $\bs y_i$. Thus, when using MCMC samplers to explore the posterior distribution of model parameters, both $\bs\Lambda$ and $\bs y_i$; $i = 1, \ldots,n$ are not marginally identifiable due to sign-switching across the MCMC trace. However, one could use the approximate maximum a posteriori estimate arising from the MCMC output in order to infer the mode of the posterior distribution of the specific parameters. These estimates correspond to the parameter values obtained at the iteration that maximizes the posterior distribution across the MCMC run. Another possibility is to restrict our attention to sign-invariant parameter functions. For example, notice that the covariance matrix $\bs\Lambda\bs\Lambda^T + \bs\Sigma$ is invariant when switching the sign of each element in $\bs\Lambda$. Following \cite{sabatti2006bayesian}, we also define the ``regularised score'' of variable $r$ to factor $j$ as 
\begin{equation}\label{eq:regExp0}
\zeta_{rj} := \lambda_{rj} \frac{\sum_{i=1}^{n}y_{ij}}{n},
\end{equation}
for $r = 1, \ldots, p$; $j =1, \ldots, q$. Notice that Equation \eqref{eq:regExp0} is invariant to simultaneously switching the signs of $\lambda_{rj}$ and $y_{ij}$'s, therefore, the estimation of $\mathbb E(\zeta_{rj}| \bs x)$ is meaningful.

\subsection{Finite Mixtures of Factor Analyzers}\label{sec:MFA}
In this section the typical factor analysis model is generalized in order to take into account unobserved heterogeneity. Assume that there are $K$ underlying groups in the population, where the number of clusters $K\geqslant 1$ denotes a known integer. Each cluster is characterized by different structure, which is reflected in our model by assuming a distinct set of parameters $(\bs\mu_k, \bs\Lambda_k, \bs\Sigma_k)$; $k = 1, \ldots,K$. Consider the latent allocation parameters $z_i\in\{1\ldots,K\}$ which assign observation $\bs x_i$ to a cluster $k =1,\ldots,K$ for $i = 1,\ldots,n$. Then, given the cluster allocations, each observation is expressed as
$$(\bs x_i|z_i = k) = \bs\mu_{k} + \bs\Lambda_{k} \bs y_i + \bs\varepsilon_i.$$
Let $0\leqslant w_k\leqslant 1$; $k = 1,\ldots,K$ and $\sum_{k=1}^{K}w_k = 1$. A-priori each observation is generated from cluster $k$ with probability equal to $w_k$, that is,
\begin{equation}\label{eq:z_prior}
\mathrm{P}(z_i = k) = w_k,\quad k = 1,\ldots,K,
\end{equation}
independent for $i = 1,\ldots,n$. We will refer to $\bs w := (w_1,\ldots,w_K)$ as the vector of mixing proportions of the model. Note that the allocation vector $\bs z := (z_1,\ldots,z_n)$ is not observed, so it should be treated as missing data.

In general, the latent factor space can be different for each cluster, e.g.~$\bs y_i|z_i = k \sim \mathcal N_{q_k}(\bs 0,\bs{\mathrm{I}}_{q_k})$, where $q_k$ represents the latent dimensionality of cluster $k$; $k = 1,\ldots,K$. Following \cite{Fokoue2003}  we assume that $(\bs y_i, z_i)$ are independent and that the intrinsic latent dimension is the same for each cluster, so the marginal distribution of $\bs y_i$ is still described by Equation \eqref{eq:y}. 

Now, we express Equations \eqref{eq:same_sigma}, \eqref{eq:x_given_y}  conditionally on the cluster membership
\begin{eqnarray}
\bs\varepsilon_i|z_i = k&\sim& \mathcal N_p(\bs 0,\bs\Sigma_{k})\label{eq:varepsilon}\\
\bs x_i|z_i = k,\bs y_i&\sim& \mathcal N_p(\bs\mu_{k}+\bs\Lambda_{k} \bs y_i,\bs\Sigma_{k}), \label{eq:fullConditionalX}
\end{eqnarray}
independent for $i = 1,\ldots,n$. Thus, Equation \eqref{eq:x_marginal} becomes
\begin{equation}
\bs x_i \sim \sum_{k = 1}^{K}w_k\mathcal N_p(\bs\mu_{k},\bs\Lambda_{k}\bs\Lambda_{k}^T + \bs\Sigma_{k}),\label{eq:mixture}
\end{equation}
independent for $i = 1,\ldots,n$. As shown in Equation \eqref{eq:mixture}, the marginal distribution of $\bs x_i$ is a finite mixture of distributions with $K$ components. Notice that when $q = 0$, the parameters $\bs\Lambda_k$, $k=1,\ldots,K$ and latent data $\bs y_i$ are no longer present in the model. Thus, Equation \eqref{eq:mixture} becomes $\bs x_i\sim\sum_{k=1}^{K}w_k\mathcal N_p(\bs\mu_k,\bs\Sigma_k)$, that is, a mixture model with diagonal variance structure per component.

A further assumption, also imposed by \cite{Fokoue2003}, is the restriction of the error variance, that is, 
\begin{equation}\label{eq:sameSigma}
\bs\Sigma_k = \bs\Sigma, \quad k = 1,\ldots,K.
\end{equation}
In this case $(z_i, \bs\varepsilon_i)$ are independent and the marginal distribution of $\bs\varepsilon_i$ is given in Equation \eqref{eq:same_sigma}. Although we consider both cases, in the following sections we present the general model where the variance of errors is allowed to vary between clusters.

Finally, the generalization of Equation \eqref{eq:regExp0} to the case that there are $K$ clusters is straightforward. For cluster $k$, the regularized score of variable $r$ to factor $j$ is defined as 
\begin{equation}\label{eq:regExp}
\zeta_{krj} := \lambda_{krj} \frac{\sum_{i=1}^{n}I(z_i = k)y_{ij}}{\sum_{i=1}^{n}I(z_i = k)},
\end{equation}
for $k = 1,\ldots, K$; $r = 1, \ldots, p$ and $j =1, \ldots, q$, where $I(\cdot)$ denotes the indicator function. Note that Equation \eqref{eq:regExp} is not defined in case where $\sum_{i=1}^{n}I(z_i = k)=0$, however this is not a problem in our implementation due to the fact that we only make inference on the ``alive'' clusters, that is, the subset of $\{1,\ldots,K\}$ defined as $\boldsymbol{K_0}=\{k=1,\ldots,K: \sum_{i=1}^{n}I(z_i=k)>0\}$.

\subsection{EM-based approaches and available software}\label{sec:previous}

A compehensive perspective on the history and development of MFA models is given in Chapter 3 of the monograph by \cite{mcnicholas2016mixture}.  \cite{ghahramani1996algorithm} applied the EM algorithm for estimating the MFA model  \eqref{eq:mixture}, under the constraint \eqref{eq:sameSigma}. The general model where $\bs\Sigma_k$ is allowed to differ between components was estimated by \cite{McLachlan:00}. \cite{tipping1999mixtures} considered the case of isotropic error variance, that is, the covariance matrix of component $k$ is written as $\bs\Lambda_k\bs\Lambda_k^T + \sigma_k \bs{\mathrm{I}}_p$ for $\sigma_k>0$ (a model which is referred to as a mixture of probabilistic principal component analyzers). 

\cite{McNicholas2008} extended the covariance structure by considering the constraints: $\bs \Lambda_k  = \bs\Lambda$, $\bs\Sigma_k=\bs\Sigma$ and $\bs\Sigma_k = \sigma_k\bs{\mathrm{I}}_p$. Furthermore, \cite{doi:10.1093/bioinformatics/btq498} introduced the parameterization $\bs\Lambda_k\bs\Lambda_k^T + \sigma_k\bs\Delta_k$, where $\sigma_k>0$ and $\bs\Delta_k = \mbox{diag}(\delta_1,\ldots,\delta_p)$ such that $|\bs\Delta_k| = 1$, with the optional constraints $\bs\Delta_k =\bs\Delta$ or $\bs\Delta_k =\bs{\mathrm{I}}_p$ for $k = 1,\ldots,K$. Depending on whether a particular constraint is present or not, a set of 12 possible models arises which is referred to as the expanded parsimonious Gaussian mixture models (EPGMM) family. A detailed description is provided in Table 2 of \cite{doi:10.1093/bioinformatics/btq498}. These models are estimated by the alternating expectation-conditional maximization (AECM) algorithm \citep{RSSB:RSSB082}.

{\tt EMMIXmfa} \citep{EMMIXmfa} is a freely available software in the form of an {\tt R} \citep{rcitation} package, implementing the approach used by \cite{McLachlan:00} and allows both constrained and uncostrained error variance per component. The {\tt pgmm} package \citep{pgmm} is available from the Comprehensive {\tt R} Archive Network and estimates MFA models using the EPGMM family \citep{McNicholas2008, doi:10.1093/bioinformatics/btq498}. Since  {\tt pgmm} offers substantially greater flexibility than {\tt EMMIXmfa}, we only report results based on the former package.

\subsection{Bayesian formulation}\label{sec:priors}

This section introduces the Bayesian framework for the MFA model, given $K$, $q$ and a random sample size of $n$ observations $\bs x = (\bs x_1,\ldots,\bs x_n)$ from Equation \eqref{eq:mixture}. Our aim is to estimate the model parameters: $(\bs w, \bs\theta)$, where $\bs \theta = \{(\bs\mu_k,\bs\Lambda_k,\bs\Sigma_k); k = 1,\ldots,K\}$. The cluster assignments as well as the latent factors $(\bs z, \bs y)$ are treated as missing data. Let $\mathcal D(\cdots)$  denote the Dirichlet distribution and $\mathcal G(\alpha,\beta)$ denote the Gamma distribution with mean $\alpha/\beta$. Let also $\bs\Lambda_{kr\cdot}$ denote the $r$-th row of the matrix of factor loadings $\bs\Lambda_k$; $k = 1,\ldots,K$; $r = 1,\ldots,p$. The following prior assumptions are imposed on the model parameters:
\begin{eqnarray}
\bs w &\sim&\mathcal D(\gamma_1,\ldots,\gamma_K) \label{eq:dirichlet_prior}\\
\bs\mu_k &\sim&\mathcal N_p(\bs\xi, \bs\Psi), \label{eq:mu_prior} \quad\mbox{independent for }k = 1,\ldots,K\\
\bs\Lambda_{kr\cdot} &\sim&\mathcal N_{\nu_r}(\bs 0,\bs\Omega), \quad\mbox{independent for }r = 1,\ldots,p \label{eq:lambda_prior} \\
\sigma_{kr}^{-2} &\sim& \mathcal G(\alpha,\beta), \quad\mbox{independent for }k = 1,\ldots,K; r = 1,\ldots,p\\
\omega_{\ell}^{-2} &\sim& \mathcal G(g,h), \quad\mbox{independent for }\ell = 1,\ldots,q \label{eq:omega_prior}
\end{eqnarray}
where all variables are assumed mutually independent and $\nu_r =\min\{r,q\}$; $r=1,\ldots,p$; $\ell = 1,\ldots,q$; $j=1,\ldots,K$.  In Equation \eqref{eq:lambda_prior}  $\bs\Omega = \mbox{diag}(\omega_1^2,\ldots,\omega_q^2)$ denotes a $q\times q$ diagonal matrix, where the diagonal entries are distributed independently according to Equation \eqref{eq:omega_prior}. The prior assumptions are the same as the ones introduced in \cite{Fokoue2003} for fixed $K$ and $q$.

\begin{figure}[t]
\centering\includegraphics[scale=0.3]{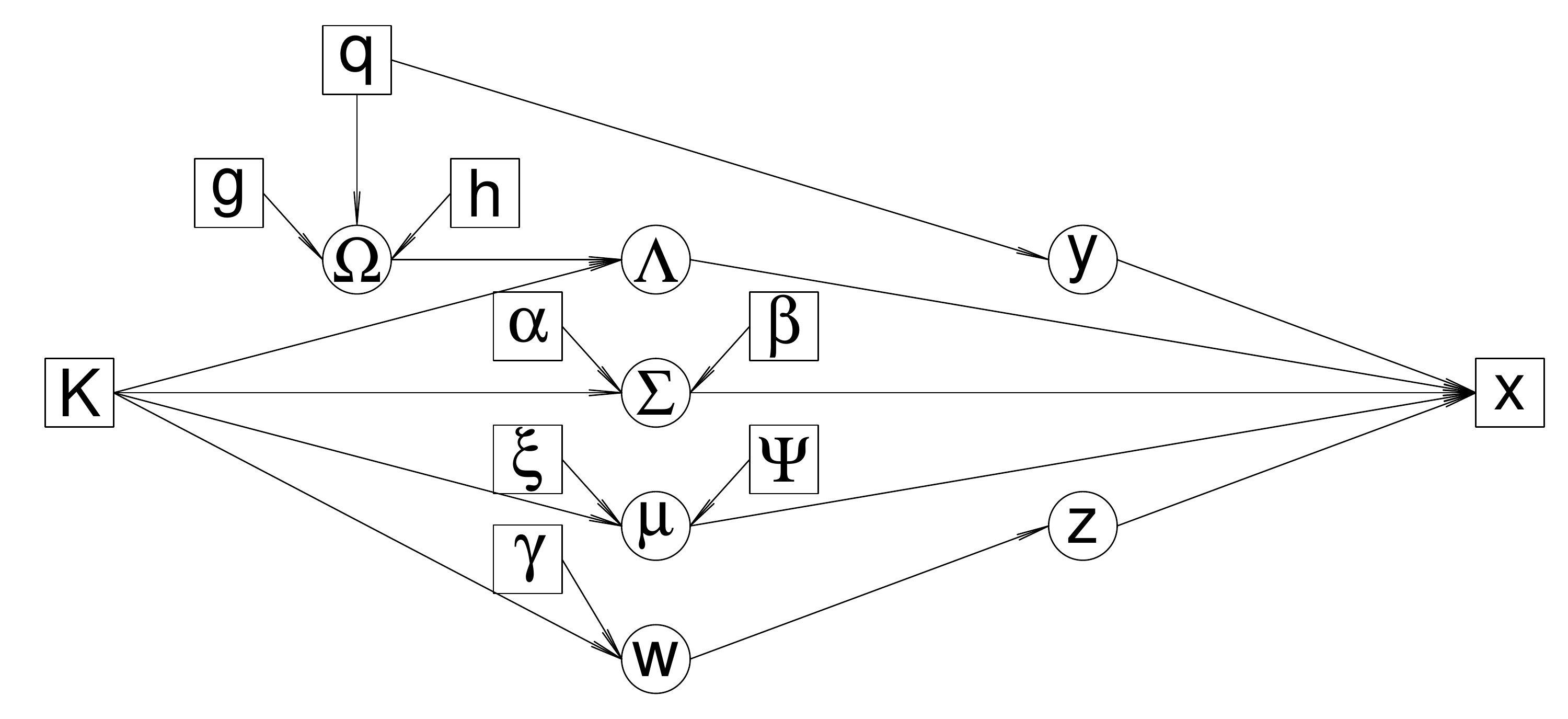}
\caption{Directed Acyclic Graph representation of the hierarchical model used in Equation \eqref{eq:joint_pdf}. Squares denote fixed/observed quantities while circles denote unknown variables.}
\label{fig1}
\end{figure}

Given the fixed set of hyperparameters $(K,q,\alpha,\beta,\bs\xi,\bs\Psi,\gamma,g,h)$, the joint probability density function of the model is written as:
\begin{align}\nonumber
f( \bs x, \bs y, \bs z,  \bs w,  \bs\mu, \bs\Sigma,\bs\Lambda, \bs\Omega) = f( \bs x| \bs y, \bs z, \bs\mu,\bs\Lambda,\bs\Sigma)f(\bs z|\bs w)f( \bs\mu)f(\bs\Sigma)\\ \times f(\bs\Lambda|\bs\Omega)
f(\bs\Omega)f( \bs w)f(\bs y)\label{eq:joint_pdf}
\end{align}
and its graphical representation is shown in Figure \ref{fig1}. In order to estimate the model parameters we use the Gibss sampler to approximately sample from the posterior distribution $f(\bs y,\bs  z,\bs  w, \bs \mu,\bs\Sigma,\bs\Lambda,\bs \Omega|\bs x)$. The reader is referred to \ref{S1_Appendix} for details.

\subsection{Overfitted mixture model}\label{sec:overfitting}

Assume that the observed data has been generated from a mixture model with $K_0$ components $$f_{K_0}(\bs x) = \sum_{k=1}^{K_0}w_kf_k(\bs x|\bs\theta_k),$$ where $f_k\in\mathcal F_{\Theta}=\{f(\cdot|\bs\theta):\bs\theta\in\Theta\}$; $k = 1,\ldots,K_0$ denotes a member of a parametric family of distributions. Consider that an overfitted mixture model $f_{K}(\bs x)$ with $K > K_0$ components is fitted to the data. It has been shown \citep{rousseau2011asymptotic} that the asymptotic behaviour of the posterior distribution of the $K-K_0$ redudant components depends on the prior distribution of mixing proportions $(\bs w)$. Let $d$ denote the dimension of free parameters of the emission distribution $f_k$. For the case of a Dirichlet prior distribution in Equation \eqref{eq:dirichlet_prior}, if $\max\{\gamma_k; k=1,\ldots,K\} < d/2$ then the posterior weight of the extra components converges to zero (Theorem 1 of \cite{rousseau2011asymptotic}). 

Let $f_K(\bs\theta,\bs z|\bs x)$ denote the joint posterior distribution of model parameters and latent allocation variables for a model with $K$ components. When using an overfitted mixture model, the inference on the number of clusters reduces to (a): choosing a sufficiently large value of mixture components ($K$), (b): running a typical MCMC sampler for drawing samples from the posterior distribution $f_K(\bs\theta, \bs z|\bs x)$ and (c) inferring the number of ``alive'' mixture components.  Note that at MCMC iteration $t = 1,2,\ldots$ (c) reduces to keeping track of the number of elements in the set $\boldsymbol{K_0}^{(t)}=\{k=1,\ldots,K: \sum_{i=1}^{n}I(z_i^{(t)}=k)>0\}$, where $z_i^{(t)}$ denotes the simulated allocation of observation $i$ at iteration $t$. At this point, we underline the fact that the MCMC sampler always operates on a space with $K$ components. The empty components at a given iteration may contain allocated observations in subsequent iterations.

In our case the dimension of free parameters in the emission distribution is equal to $d = 2p + pq-\frac{q(q-1)}{2}$. We set $\gamma_1=\ldots=\gamma_K = \frac{\gamma}{K}$, thus the distribution of mixing proportions in Equation \eqref{eq:dirichlet_prior} becomes
\begin{equation}\label{eq:dirichlet_prior_same}
 \bs w\sim\mathcal D\left(\frac{\gamma}{K},\ldots,\frac{\gamma}{K}\right)
\end{equation}
where $0 < \gamma < d/2$ denotes a pre-specified positive number. Such a value is chosen for two reasons. At first, it is smaller than $d/2$ so the asymptotic results of \cite{rousseau2011asymptotic} ensure that extra components will be emptied as $n\rightarrow\infty$. Second, this choice can be related to standard practice when using Bayesian non-parametric clustering methods where the parameters of a mixture are drawn from a Dirichlet process \citep{ferguson}, that is, a Dirichlet process mixture model \citep{neal2000markov}. This approach is used in a variety of applications that involve clustering of subjects into an unknown number of groups, see for example \cite{medvedovic2002bayesian} where they cluster gene expression profiles using infinite mixture models. Under this point of view, the prior distribution in Equation \eqref{eq:dirichlet_prior_same} represents a finite-sum random probability measure that approximates the Dirichlet process with concentration parameter $\gamma$ \citep{ishwaran2002exact}.

\subsection{Selecting the Number of Factors}\label{sec:qSelection}

Recall that the MFA model has been defined assuming that the number of factors $q$ is fixed. However, the dimensionality of the latent factor space is not known and should be estimated. In \cite{Fokoue2003} the number of factors is treated as a random variable and a Birth-Death MCMC sampler \cite{stephens2000} is used to draw samples from the joint posterior distribution of model parameters and $q$. Treating $q$ as a random variable is preferable from a Bayesian point of view, however it is noted that the sampler may exhibit poor mixing, especially in high dimensions. In our setup, we consider a simpler approach where we produce MCMC samples conditionally on each value of $q$ in a pre-specified range and choose the best one according to model selection criteria. 

The following penalized likelihood criteria were used: Akaike's information Criterion (AIC) \citep{akaike}, Bayesian Information Criterion (BIC) \citep{Schwarz:78}, and Deviance Information Criterion (DIC) \citep{spiegelhalter2002bayesian}.  
Let $D(\theta) = -2\log f( x; \theta)$ denote the deviance. Then, 
\begin{eqnarray*}
\mbox{AIC}(K_0,q) &=& D( \widehat{\bs\theta}) + 2d_{K_0,q}\\
\mbox{BIC}(K_0,q) &=& D(\widehat{\bs\theta}) + d_{K_0,q}\log n\\
\mbox{DIC}(K_0,q) & = & D( \widehat{\bs\theta}) +2 p_D\\
&=&-4\mathbb{E}_{\bs \theta}\left[\log\{f_{K_0}(\bs x; \bs \theta)\}\right] + 2 \log\{f_{K_0}(\bs x;  \widehat{\bs\theta})\},
\end{eqnarray*}
where $d_{K_0,q}$ is the number of free parameters of a model with $K_0$ clusters and $q$ factors, whereas $p_D = \overline{D(\bs\theta)} - D(\widehat{\bs\theta})$ denotes the \textit{effective dimension} of the model. It is well-known that DIC tends to overfit \citep{spiegelhalter2014deviance}, so we also consider an alternative version which increases the penalty term, namely $\mbox{DIC}_2 = D( \widehat{\bs\theta}) +3 p_D$ \citep{linde2005dic, van2012bayesian}. For AIC and BIC, $\widehat{\bs\theta}$ corresponds to the Maximum Likelihood estimate of $\bs\theta$. Note that under certain conditions \citep{doi:10.1080/01621459.1995.10476592}, the quantity $-\mbox{BIC}/2$ achieves asymptotic consistency by roughly approximating the logarithm of the Bayes factor \citep{kassRaftery} between two models. The definition of $\widehat{\bs\theta}$ is not unique for DIC (see the discussion in \cite{celeux2006deviance}). We have chosen $\widehat{\bs\theta}$ as the simulated parameter values that maximizes the observed log-likelihood across the MCMC run. 

It should be noted that in order to compute the observed likelihood as well as the number of free parameters $d_{K_0,q}$ in AIC and BIC we used only the parameter values of ``alive'' components, corresponding to the most probable number of clusters (after rescaling the weights in order to sum to 1). It does not make sense to include the redundant parameters in the computation of the number of parameters simply because the results will be inconsistent when considering various values for the total number of clusters in the overfitted mixture ($K$). Also recall here that asymptotically the redundant components are assigned zero posterior weight. This means that the contribution of the extra components to the observed log-likelihood is zero. 

\subsection{Prior Parallel Tempering}\label{sec:tempering}

It is well known that the posterior surface of mixture models can exhibit many local modes \citep{celeux2000computational, Marin:05}. In such cases simple MCMC algorithms may become trapped in minor modes and demand a very large number of iterations to sufficiently explore the posterior distribution. In order to produce a well-mixing MCMC sample and improve the convergence of our algorithm we utilize ideas from parallel tempering schemes \citep{geyer1991, geyer1995, Altekar12022004}, where different chains are running in parallel and they are allowed to switch states. Each chain corresponds to a different posterior distribution, and usually each one represents a ``heated'' version of the target posterior distribution. This is achieved by raising the original target to a power $T$ with $0\leqslant T \leqslant 1$, which flattens the posterior surface, thus, easier to explore when using an MCMC sampler. 

In the context of overfitting mixture models, \cite{overfitting} introduced a prior parallel tempering scheme. Under this approach, each heated chain corresponds to a model with identical likelihood as the original, but with a different prior distribution. Although the prior tempering can be imposed on any subset of parameters, it is only applied to the Dirichlet prior distribution of mixing proportions. Let us denote by $f_i(\bs\varphi|\bs x)$ and $f_i(\bs\varphi)$; $i=1,\ldots,J$, the posterior and prior distribution of the $i$-th chain, respectively. Obviously, $f_i(\bs\varphi|\bs x) \propto f(\bs x|\bs\varphi)f_i(\bs\varphi)$. Let $\bs\varphi^{(t)}_i$ denote the state of chain $i$ at iteration $t$ and assume that a swap between chains $i$ and $j$ is proposed. The proposed move is accepted with probability $\min\{1,A\}$ where
\begin{equation}\label{eq:mh_ar}A = \frac{f_i(\bs\varphi_j^{(t)}|\bs x)f_j(\bs\varphi_i^{(t)}|\bs x)}{f_i(\bs\varphi_i^{(t)}|\bs x)f_j(\bs\varphi_j^{(t)}|\bs x)}=
\frac{f_i(\bs\varphi_j^{(t)})f_j(\bs\varphi_i^{(t)})}{f_i(\bs\varphi_i^{(t)})f_j(\bs\varphi_j^{(t)})}=
\frac{\widetilde{f}_i( w_j^{(t)})\widetilde{f}_j( w_i^{(t)})}{\widetilde{f}_i( w_i^{(t)})\widetilde{f}_j( w_j^{(t)})},\end{equation}
and $\widetilde{f}_i(\cdot)$ corresponds to the probability density function of the Dirichlet prior distribution related to chain $i = 1,\ldots,J$. According to Equation \eqref{eq:dirichlet_prior_same}, this is
\begin{equation}\label{eq:dir_prior_tempering}
 \bs w\sim\mathcal D\left(\frac{\gamma_{(j)}}{K},\ldots,\frac{\gamma_{(j)}}{K}\right),
\end{equation} 
for a pre-specified set of parameters $\gamma_{(j)}>0$ for $j = 1,\ldots,J$.

\subsection{Label Switching Problem}\label{sec:label}

The likelihood of a mixture model with $K$ components is invariant with respect to permutations of the parameters $(w_1,\theta_1),\ldots,(w_K,\theta_K)$. Typically, the same invariance property holds for the prior distribution of $(\bs w,\bs\theta)$. Therefore, the posterior distribution $f(\bs w,\bs\theta|\bs x)$ will be invariant to permutations of the labels and it will exhibit (a multiple of) $K!$ symmetric modes. The label switching problem \citep{redner1984mixture, Jasra, Papastamoulis2013} refers to the fact that an MCMC sample that has sufficiently explored the posterior surface will be switching among those symmetric areas. 

Note that the symmetry of the likelihood is not of great practical importance under a frequentist point of view, because the EM algorithm will converge to a mode of the likelihood surface. However, under a Bayesian approach it burdens the estimation procedure since the marginal posterior  distribution of $(w_k,\theta_k)$ will be the same for all $k = 1,\ldots,K$. In order to derive meaningful estimates of the marginal posterior distributions as well as estimates of the posterior means the inference should take into account the label switching problem \citep{stephens2000dealing, Marin:05, Papastamoulis:10, rodriguez, yao2014online}. A variety of relabelling algorithms is available in the {\tt label.switching} package \citep{papastamoulis2016label}, more specifically we have used the ECR reordering algorithm \citep{Papastamoulis:10}. 

\begin{figure}[t]
\centering
\begin{tabular}{cc}
\includegraphics[scale=0.25]{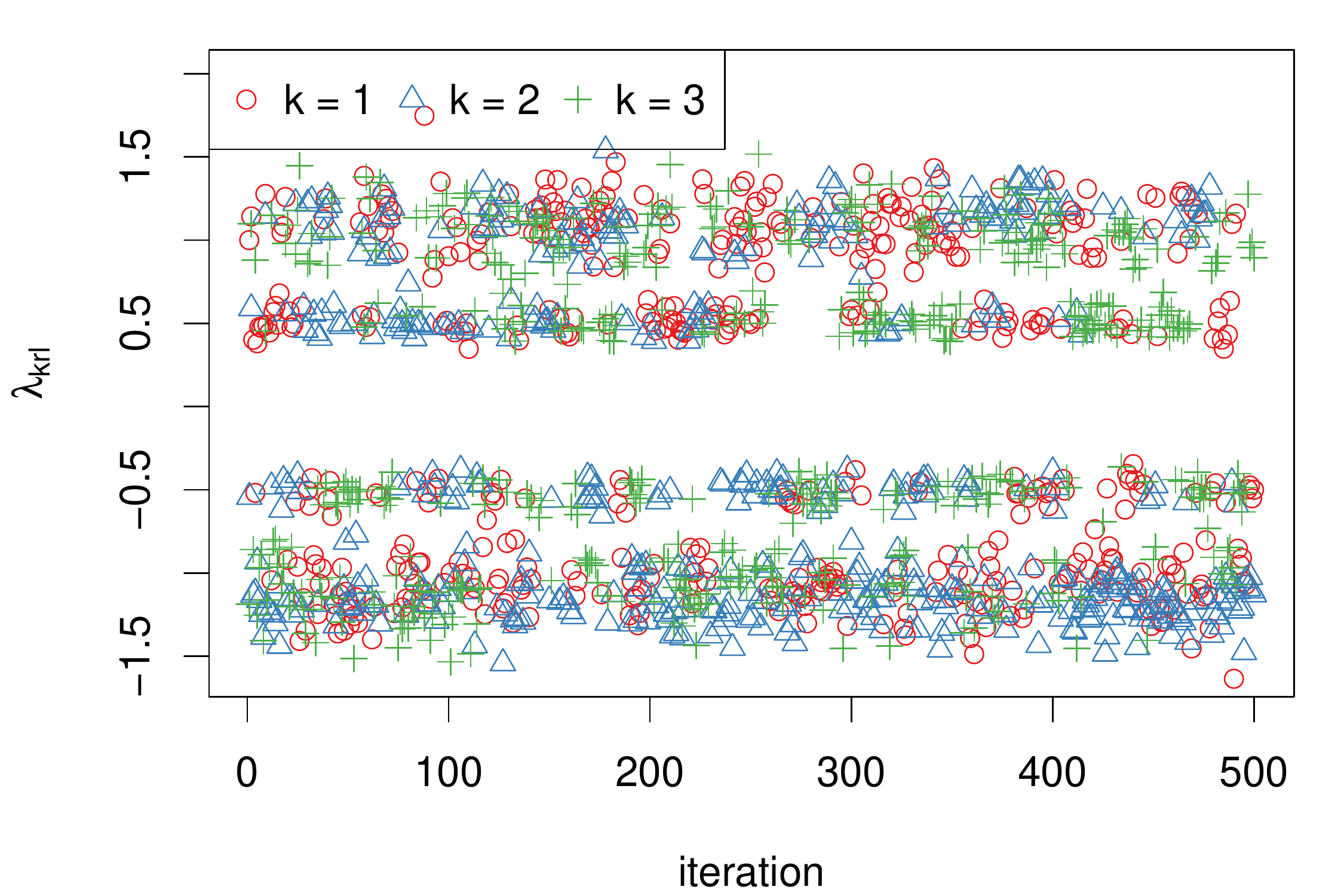} &
\includegraphics[scale=0.25]{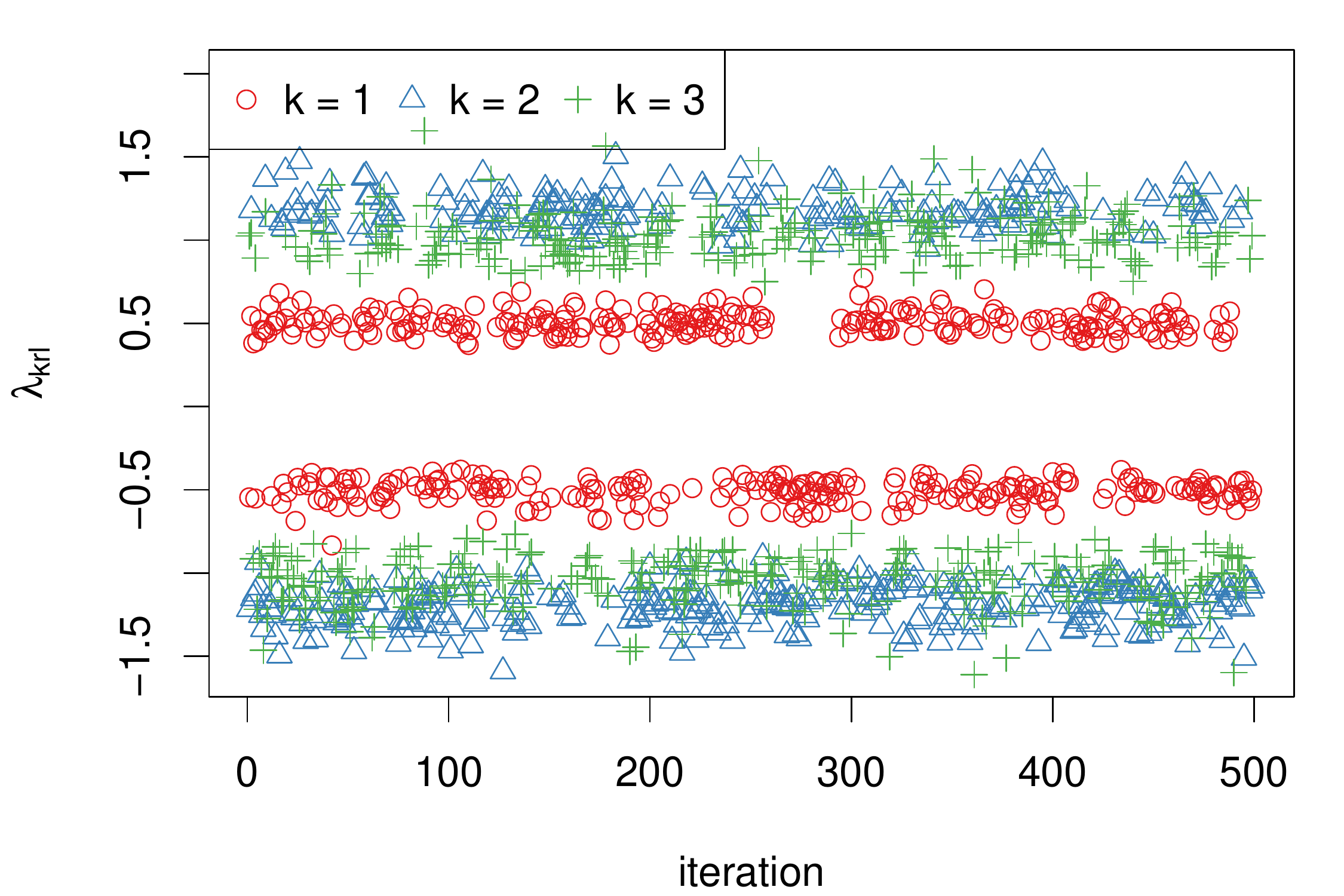}\\
(a) & (b)
\end{tabular}
\caption{Simulated factor loadings $\{\lambda_{kr\ell}; k = 1,2,3\}$, for variable $r = 1$ and latent factor $\ell = 1$, using one of the synthetic datasets described in  \ref{S2_Appendix} considering that the total number of mixture components is fixed at its true value $K = 3$. The generated values correspond to a thinned MCMC sample of 500 iterations after burn-in period. (a): MCMC trace with label switching and sign switching. (b) Reordered MCMC trace with sign-switching after eliminating the label switching via the ECR algorithm. 
}
\label{fig:switchings}
\end{figure}

Finally we underline that reordering the MCMC sample will only deal with the label switching problem but not the sign switching problem related to the rows of $\bs\Lambda_k$; $k = 1,\ldots,K$, as shown in Figure \ref{fig:switchings}. Figure \ref{fig:switchings}.(a) illustrates both label and sign switching. After successfully dealing with label switching, the MCMC trace of factor loadings shown in Figure \ref{fig:switchings}.(b) is identifiable up to a sign indeterminacy.  But as mentioned earlier, for this particular set of parameters we restrict our attention to sign-invariant parametric functions. 

\subsection{Details of the implementation}\label{sec:details}

\textbf{Number of overfitted mixture components} In our applications we have considered overfitted mixtures with $K = 20$ components. In all cases the sampled values of the number of ``alive'' components was  strictly smaller than this upper bound, at least for models where the number of factors was equal to or larger than the ``true'' one. If this is not the case for any of the fitted models, the user should consider increasing this upper bound.

\textbf{Data normalization and prior parameters} Before running the sampler, the raw data is standardized by applying the $z$-transformation
\begin{equation*}
\frac{x_{ir} - \bar x_{r}}{\sqrt{s^2_r}}, \quad i =1,\ldots,n;r = 1,\ldots,p
\end{equation*}
where $\bar x_{r} = \frac{\sum_{i=1}^{n}x_{ir}}{n}$ and $s^{2}_{r}= \frac{1}{n-1}\sum_{i=1}^{n}\left(x_{ir}-\bar x_{r}\right)^2$. The main reason for using standardized data is that the sampler mixes better. Furthermore, it is easier to choose prior parameters that are not depending on the observed data, that is, using the data twice. In any other case, one could use empirical prior distributions as reported in \cite{Fokoue2003}, see also \cite{dellaportas2006multivariate}. For example, a sensible choice for selecting the prior parameter $\boldsymbol\xi$ in case of non-standardized data is to set it equal to the vector containing the mean of each variable. The parameters $\alpha$ and $\beta$ could be selected in a way so that the mode of the prior distribution of $\boldsymbol \sigma^2_{kr}$ is equal to the variance of each variable, with a large standard deviation. However, in cases where the data contains variables which are measured in very different scales, the algorithm may be quite sensitive to the selection of these prior parameters. Therefore, we advise to use standardized data at the cost of losing interpretability of the parameter estimates.

For the case of standardized data, the prior parameters are specified in Table \ref{tab:prior}.  

\begin{table}[!ht]
\centering
\begin{tabular}{llllllll}
\toprule
 & $\alpha$& $\beta$& $\gamma$& $g$& $h$& $\bs\xi = (\xi_1,\ldots,\xi_p)^T$& $\bs\Psi$\\
  \midrule
value & 0.5  & 0.5 & 1 &  0.5 & 0.5 & $(0,\ldots,0)^T$ &  $\bs{\mathrm{I}}_p$ \\ \bottomrule
\end{tabular}
\caption{Prior parameter specification for the case of standardized data.}
\label{tab:prior}
\end{table}

\textbf{Prior parallel tempering} We used a total of $J = 8$ parallel chains where the prior distribution of mixing proportions for chain $j$ in Equation \eqref{eq:dir_prior_tempering} is selected as
\begin{equation*}
\gamma_{(j)} = \gamma + \delta(j-1), \quad j = 1,\ldots,J,
\end{equation*}
where $\delta > 0$.  Since $\gamma = 1$ (as shown in Table \ref{tab:prior}) and $K=20$, it follows from Equation \eqref{eq:dirichlet_prior_same} that the parameter vector of the Dirichlet prior of mixture weights which corresponds to the target posterior distribution ($j = 1$) is equal to $(0.05,\ldots,0.05)$. Also in our examples we have used $0.25 \leqslant \delta \leqslant 5$, but in general we strongly suggest to tune this parameter until a reasonable acceptance rate is achieved. Each chain runs in parallel and every 10 iterations we randomly select two adjacent chains $(j,j+1)$, $j\in\{1,\ldots,J-1\}$ and propose to swap their current states. A proposed swap is accepted with probability $A$ in Equation \eqref{eq:mh_ar}.

\textbf{Initialization strategy} The sampler may require a large number of iterations to discover the high posterior probability areas when initialized from totally random starting values. Of course this behaviour is alleviated as the number of parallel chains ($J$) increases, for example \cite{overfitting} obtained good mixing when $J = 30$ for mixtures of univariate normal distributions. In our case, the number of parameters is dramatically larger and an even larger number of parallel chains would be required to obtain similar levels of mixing. However we would rather to test our method in cases where the number of available cores takes smaller values, e.g.~$J=8$ or even $J = 4$, which are typical numbers of threads in modern-day laptops. Under this point of view, the following two-stage initialization scheme is adopted. 

We used an initial period of 100 MCMC iterations where each chain is initialized from totally random starting values, but under a Dirichlet prior distribution  with large prior parameter values. These values were chosen in a way that the asymptotic results of \cite{rousseau2011asymptotic} guarantee that the redundant mixture components will have non-negligible posterior weights. More specifically for chain $j$ we assume $ w\sim\mathcal D(\gamma'_{j},\ldots,\gamma'_{j})$ with $\gamma'_{(j)} = \frac{d}{2} + (j-1)\frac{d}{2(J-1)}$, for $j =1,\ldots,J$. We observed that such an initialization quickly reaches to a state where the true clusters are split into many subgroups. Then, we initialize the actual model by this state. In this case, the sampler will spend some time combining the split clusters to homogeneous groups. As soon as all similar subgroups are merged into homogeneous clusters, the sampler will start exploring the stationary distribution of the chain. A comparison with a simpler random starting scheme is presented in  \ref{sec:convergence} (see also \ref{sec:standardBMFA}).  

\textbf{Number of parallel chains, MCMC iterations and post-processing} Our MCMC sampler ran for $20000$ iterations using $8$ heated chains that run on parallel using the same number of threads. The first $5000$ iterations were discarded as burn-in period and then each chain was thinned by keeping every $10^{\mbox{th}}$ iteration. The MCMC sample corresponding to the retained iterations was reordered according to the ECR algorithm \citep{Papastamoulis:10} available in the {\tt label.switching} package \citep{papastamoulis2016label}.

\section{Results}\label{sec:results}

At first, we use an extended simulation study in order to evaluate the ability of our method to estimate the correct clustering structure. Our results are benchmarked in terms of clustering estimation accuracy against state-of-the-art software for fitting MFA models with the EM algorithm as implemented in the R package {\tt pgmm} (version 1.2) \citep{pgmm}.  Finally, we illustrate the applicability of our method to three publicly available datasets with known ground-truth classifications. In what follows, we use the acronym {\tt fabMix} to refer to the proposed method.

\subsection{Simulation study}\label{sec:simulationStudy}\label{sec:sim}

\begin{figure}[t]
\begin{tabular}{ccc}
\hspace{-3ex}\includegraphics[scale=0.4]{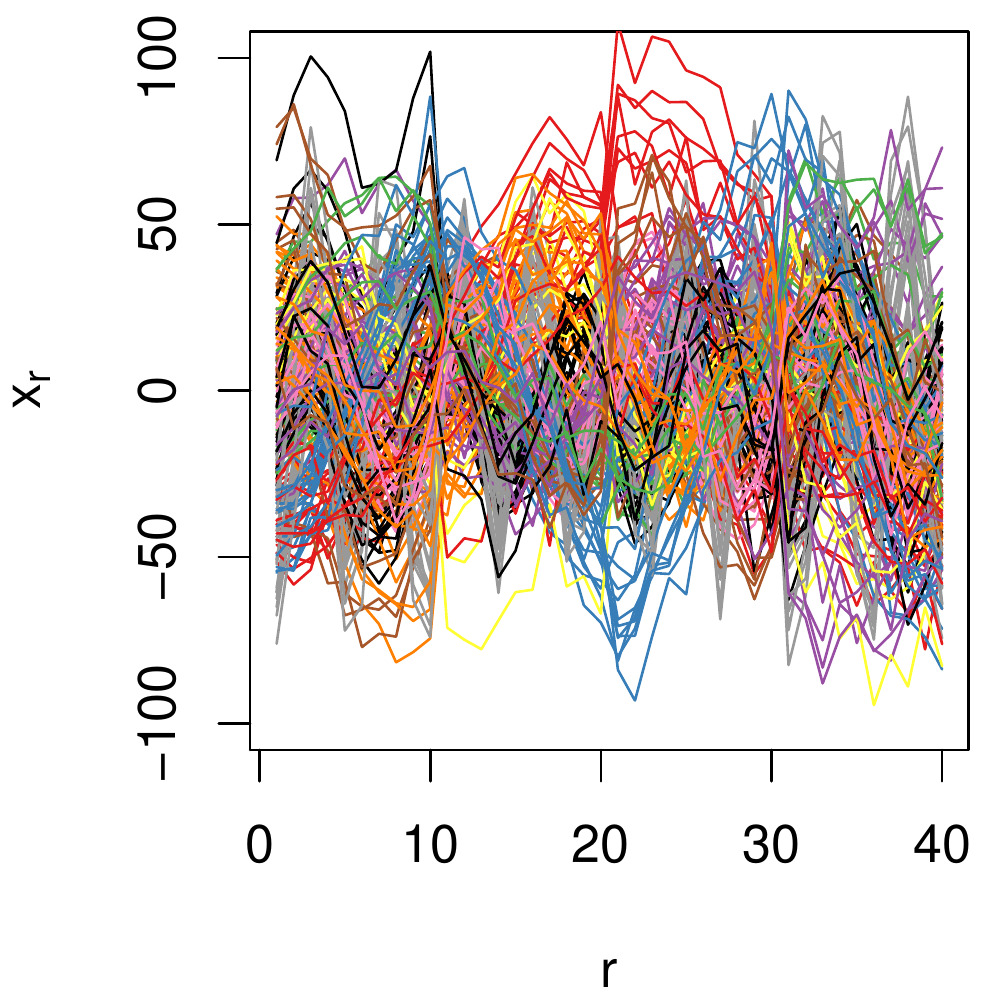} &
\hspace{-3ex}\includegraphics[scale=0.4]{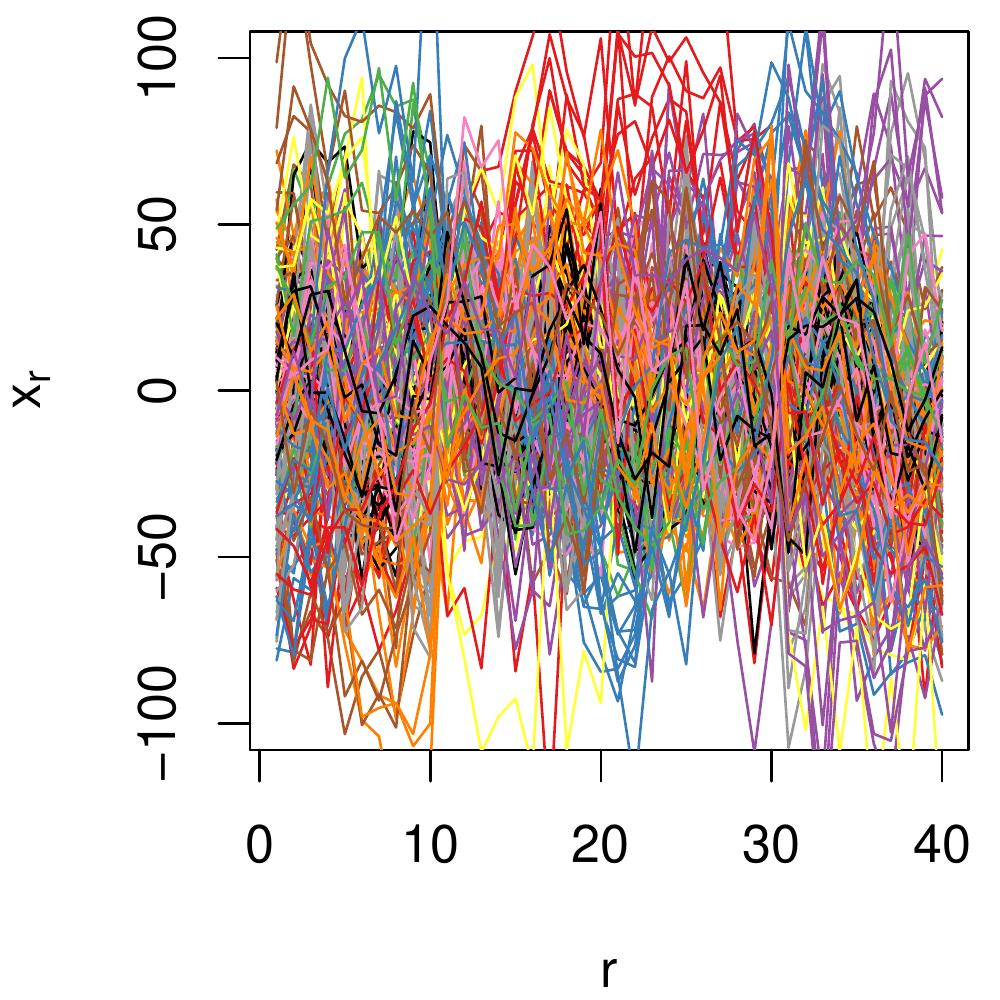} & 
\hspace{-3ex}\includegraphics[scale=0.4]{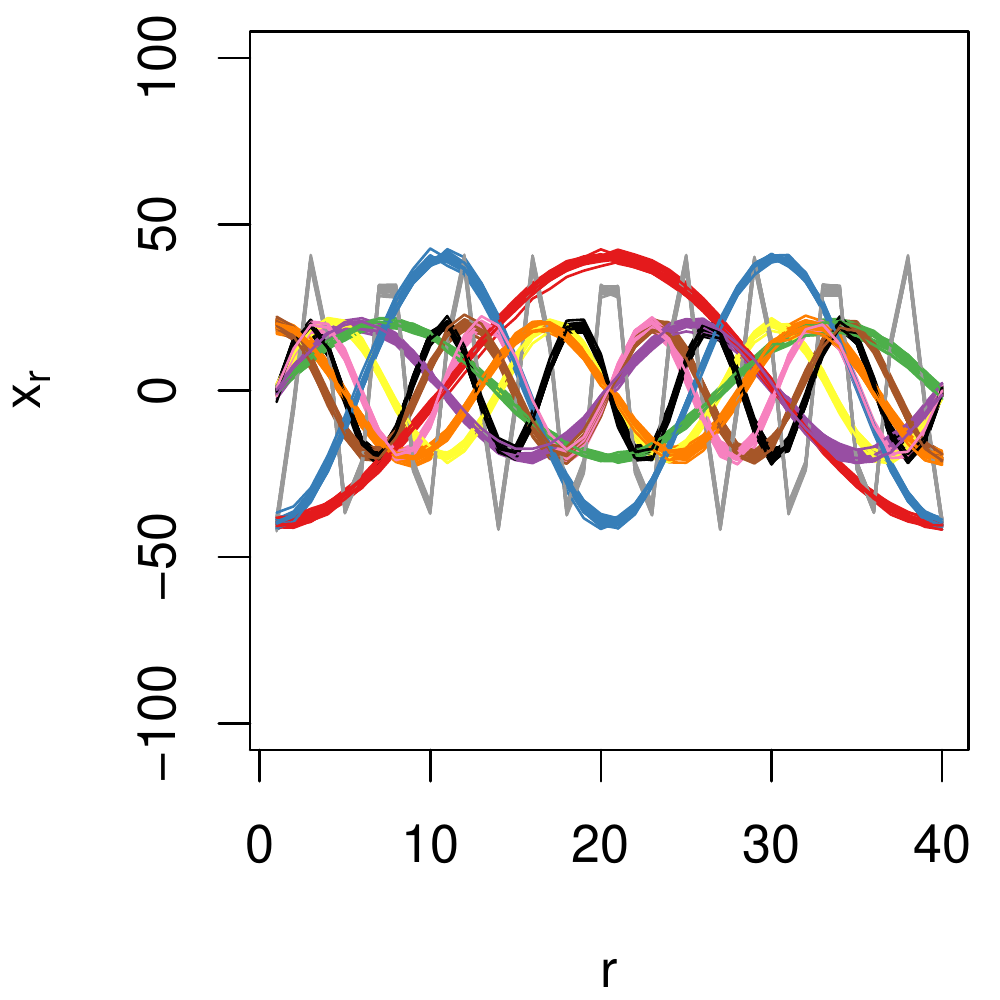}\\
\hspace{-3ex}\includegraphics[scale=0.4]{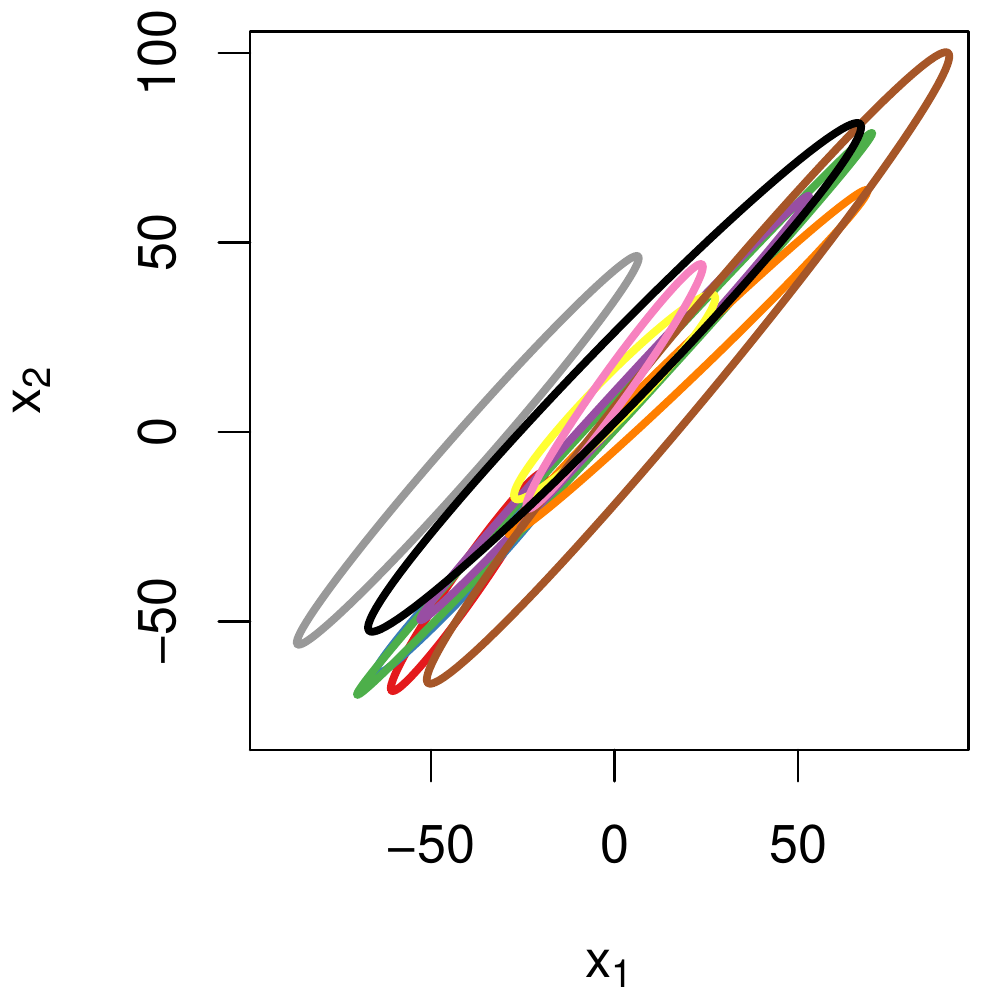} &
\hspace{-3ex}\includegraphics[scale=0.4]{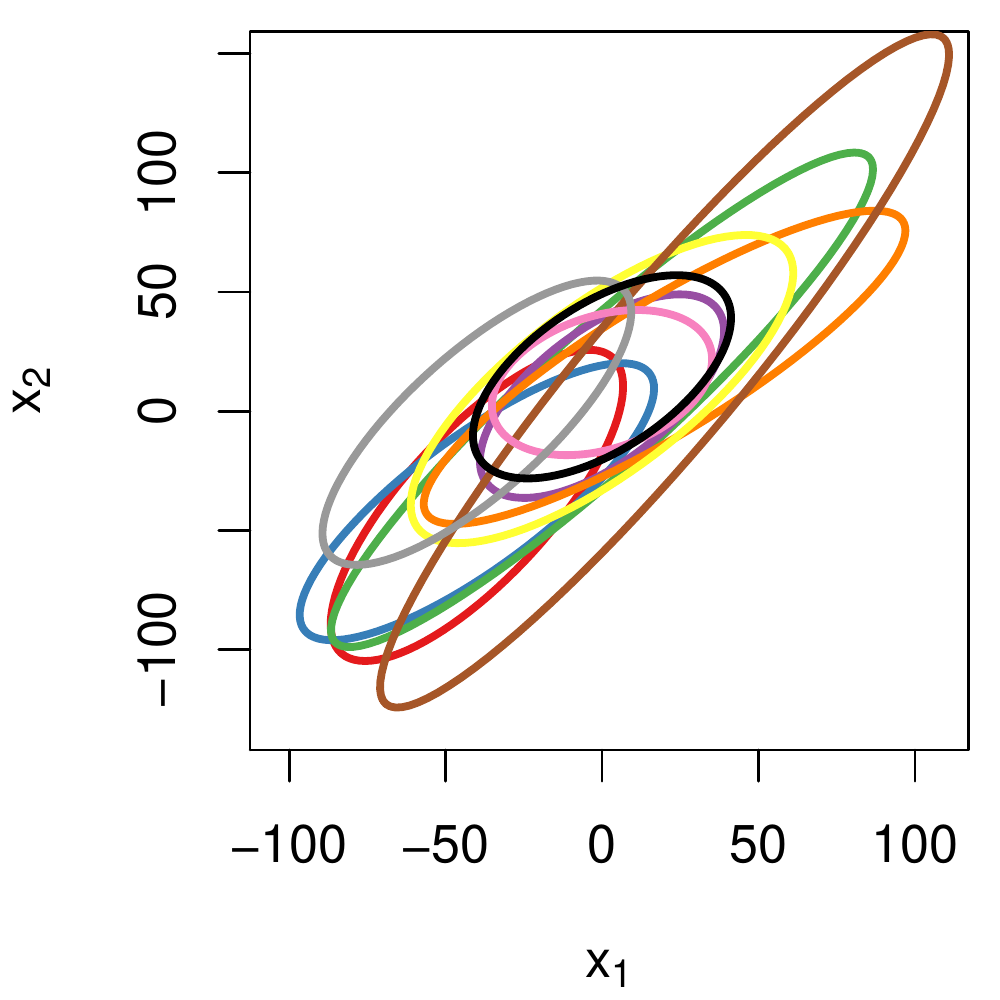} & 
\hspace{-3ex}\includegraphics[scale=0.4]{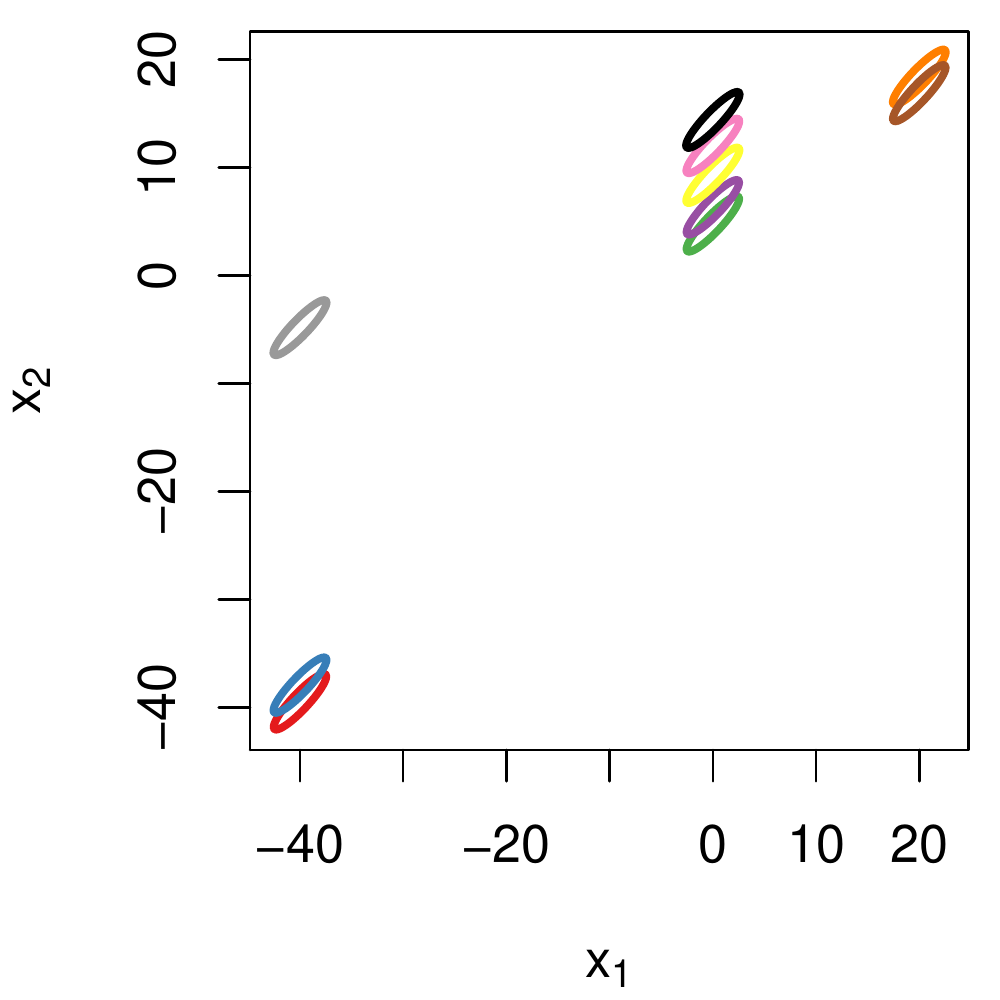}\\
(a) scenario 1  & (b) scenario 2 & (c) scenario 3
\end{tabular}
\caption{(a): A simulated dataset with $K=10$ clusters, using the generation procedure described in \ref{S2_Appendix} for $p = 40$ variables. Each color corresponds to a distint cluster. (b) and (c): Simulated datasets with the same marginal means and cluster assignments as in (a), but different covariance structures. The ellipses at the lower panel display the $95\%$ confidence region of the marginal bivariate normal distribution of $(x_1,x_2)$ per cluster centered on $(\mu_{k1},\mu_{k2})$, $k=1,\ldots,10$.
}
\label{fig:simDataExample}
\end{figure}

Figure \ref{fig:simDataExample}.(a) visualizes a simulated dataset with 10 clusters using the procedure in \ref{S2_Appendix} for $p = 40$ variables. The sample size is set to $n = 500$ (only a randomly sampled subset of 200 observations is actually displayed in Figure \ref{fig:simDataExample}). On panels (b) and (c), we juxtapose two similar datasets with identical cluster labels and marginal means, but different covariance structure. It is evident that the clusters exhibit varying levels of overlapping between different scenarios. The underlying latent factor space consists of $q = 4$ dimensions for the first two scenarios and $q = 1$ for scenario 3. 

Both methods estimated MFA models with a number of factors between $q  = 1,\ldots,8$. The EM-based method fitted a series of mixture models with $K\in\{1,\ldots,20\}$ and selected the optimal number of clusters and factors using the BIC. We also used BIC in order to choose the number of factors in our Bayesian MFA model. The reader is referred to \ref{sec:qEstimation} for a detailed comparison between BIC, AIC and DIC for selecting the number of factors in our model.

It is well known that likelihood methods may be highly sensitive to starting values and that it is quite challenging to adopt optimal schemes for the initialization of the EM algorithm in mixture models (see for example \cite{papastamoulis2016estimation}). As with all EM-type algorithms, {\tt pgmm} may converge to local maxima. We used six starting values: five runs were initialized from random starting values and one run was initialized using the $k$-means clustering algorithm. Regarding the parameterization we have considered all 12 models available in {\tt pgmm} and the best one is selected according to BIC. The wallclock run-time for fitting all 12 {\tt pgmm} models with six different starts for $1\leqslant K\leqslant 20$ and $1\leqslant q\leqslant 8$ for Scenario 1 is 62 hours, using 12 parallel threads (one thread per model). The corresponding computing time for fitting the two parameterizations of the overfitted MFA with $K = 20$ components for the same range of factors and $20000$ MCMC iterations was equal to 71 hours, based on 8 cores (one core per heated chain).

\begin{table}[t]
\centering
\begin{tabular}{cllllllll}
\toprule
& &  & \multicolumn{3}{c}{{\tt fabMix}}
  & \multicolumn{3}{c}{{\tt pgmm} }
\\ 
& $K$ & $q$  & ARI & $\widehat{K}$ & $\widehat{q}$ & ARI & $\widehat{K}$ & $\widehat{q}$\\
  \midrule
scenario 1& 10 & 4  &1 & 10 & 4 & 0.968 & 11 & 4 \\ 
scenario 2& 10 & 4 & 1 & 10 & 4 & 0.903 & 9 & 4 \\ 
scenario 3& 10 & 1 & 1 & 10 & 1 & 1 & 10 & 1  \\ \bottomrule
\end{tabular}
\caption{Adjusted Rand Index (ARI) between the estimated and true cluster assignments, estimate of the number of clusters ($\widehat{K}$) and factors ($\widehat{q}$) for the simulated datasets generated according to three scenarios with different levels of overlapping and covariance structure. }
\label{tab:overlapping}
\end{table}

The results are summarized in Table \ref{tab:overlapping}, in terms of the estimation of the number of clusters, factors and clustering accuracy based on the Adjusted Rand Index (ARI) \citep{doi:10.1080/01621459.1971.10482356}. Note that {\tt fabMix} exhibits excellent performance in all  cases. The best models selected from {\tt pgmm} correspond to the parameterizations: {\tt "UCU"}, (scenarios 1 and 2) and {\tt "CCU"} (scenario 3). In all cases, the model with the same variance of errors is selected from {\tt fabMix}. 


\begin{figure}[p]
\begin{tabular}{cc}
\hspace{-8ex}\includegraphics[scale=0.3]{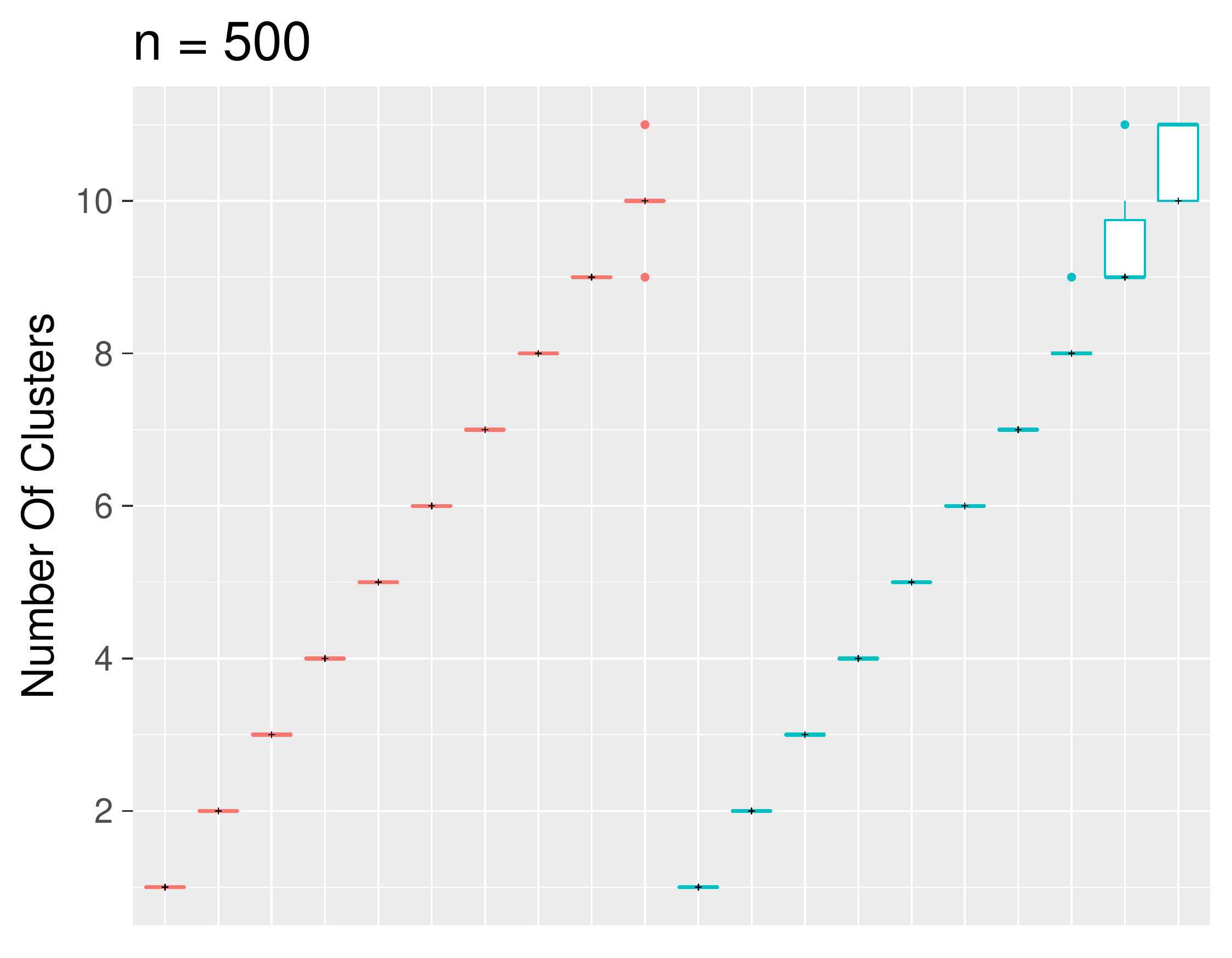} & \hspace{-3ex}\includegraphics[scale=0.3]{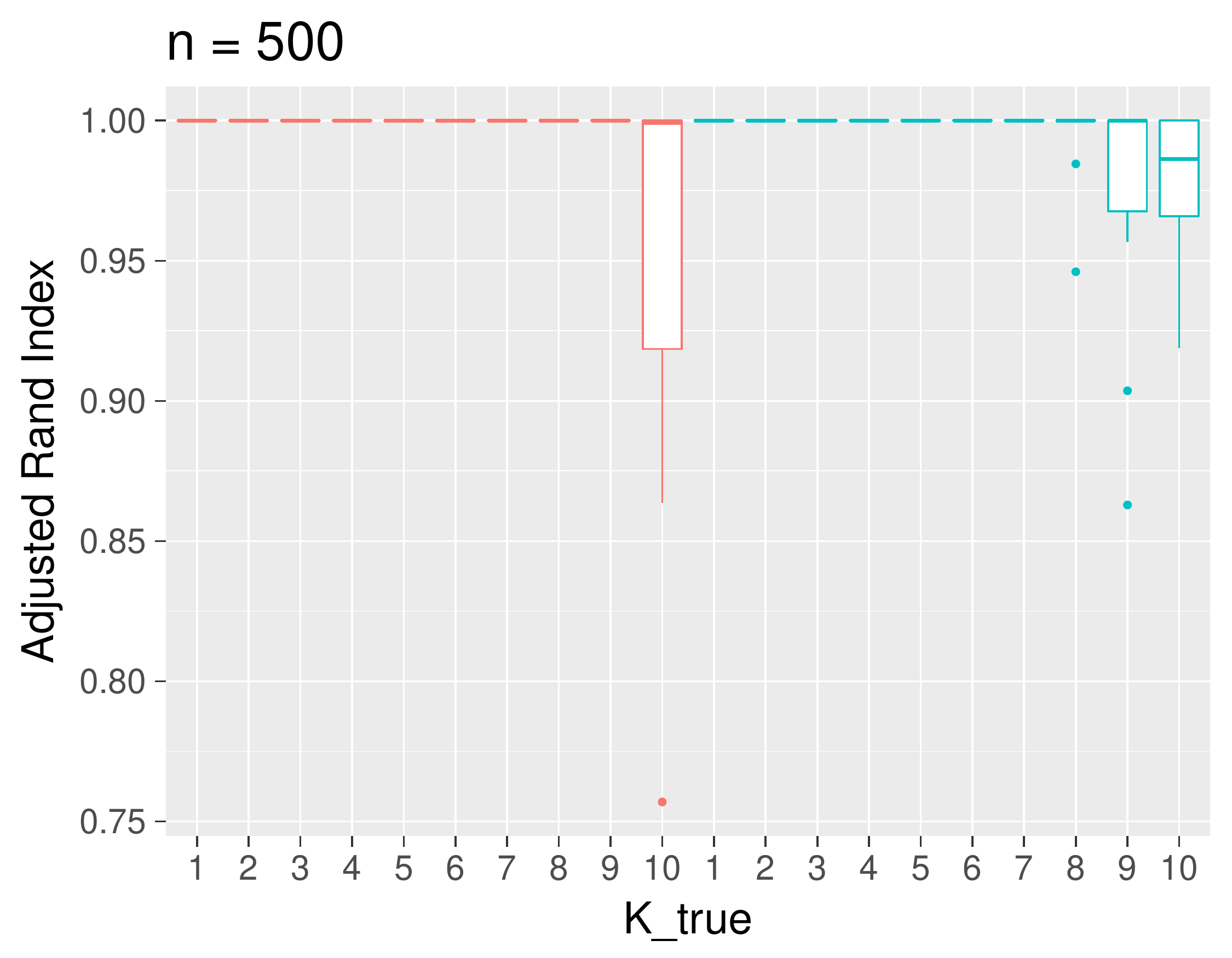} \\
\hspace{-8ex}\includegraphics[scale=0.3]{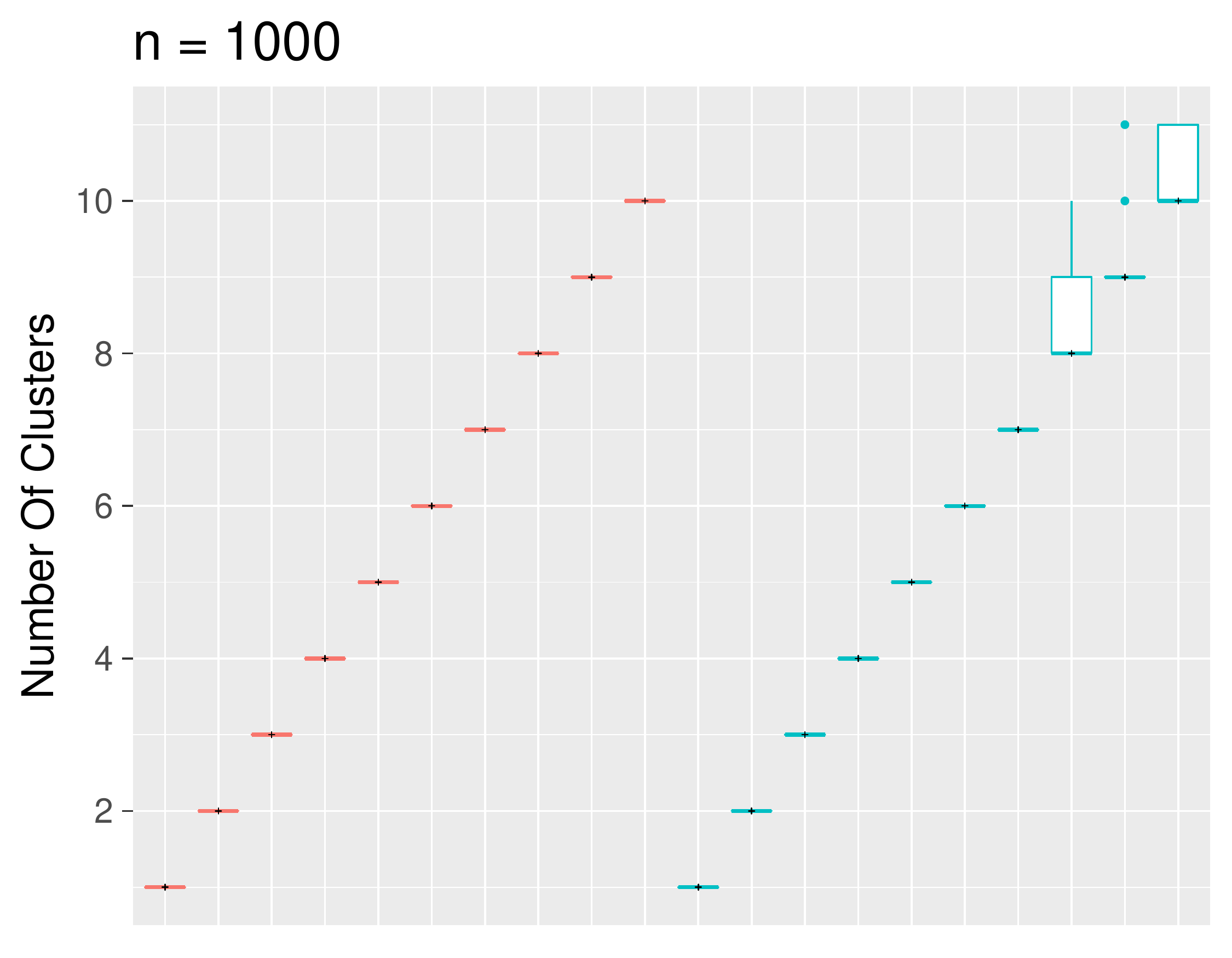} & \hspace{-3ex}\includegraphics[scale=0.3]{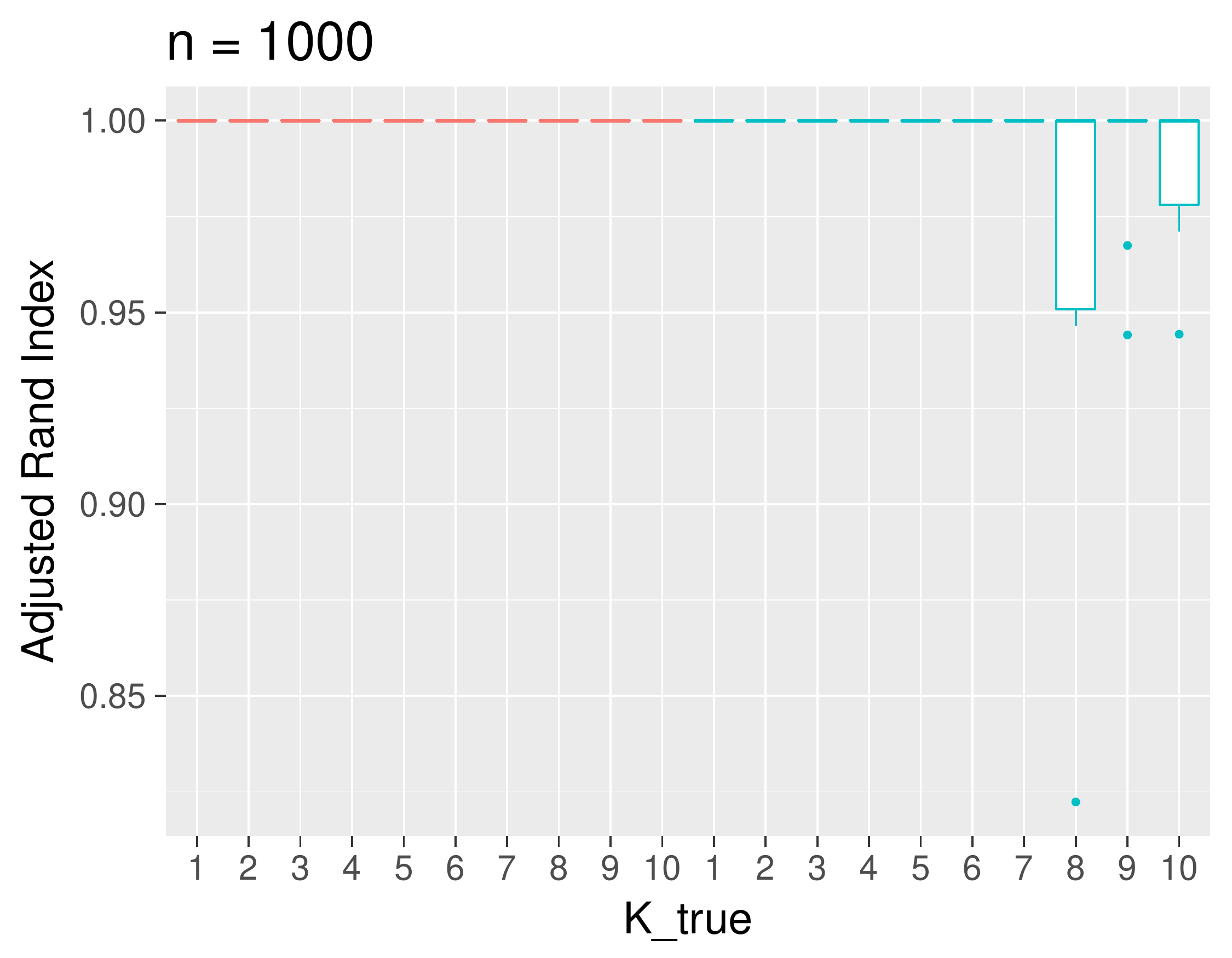} \\
\hspace{-8ex}\includegraphics[scale=0.3]{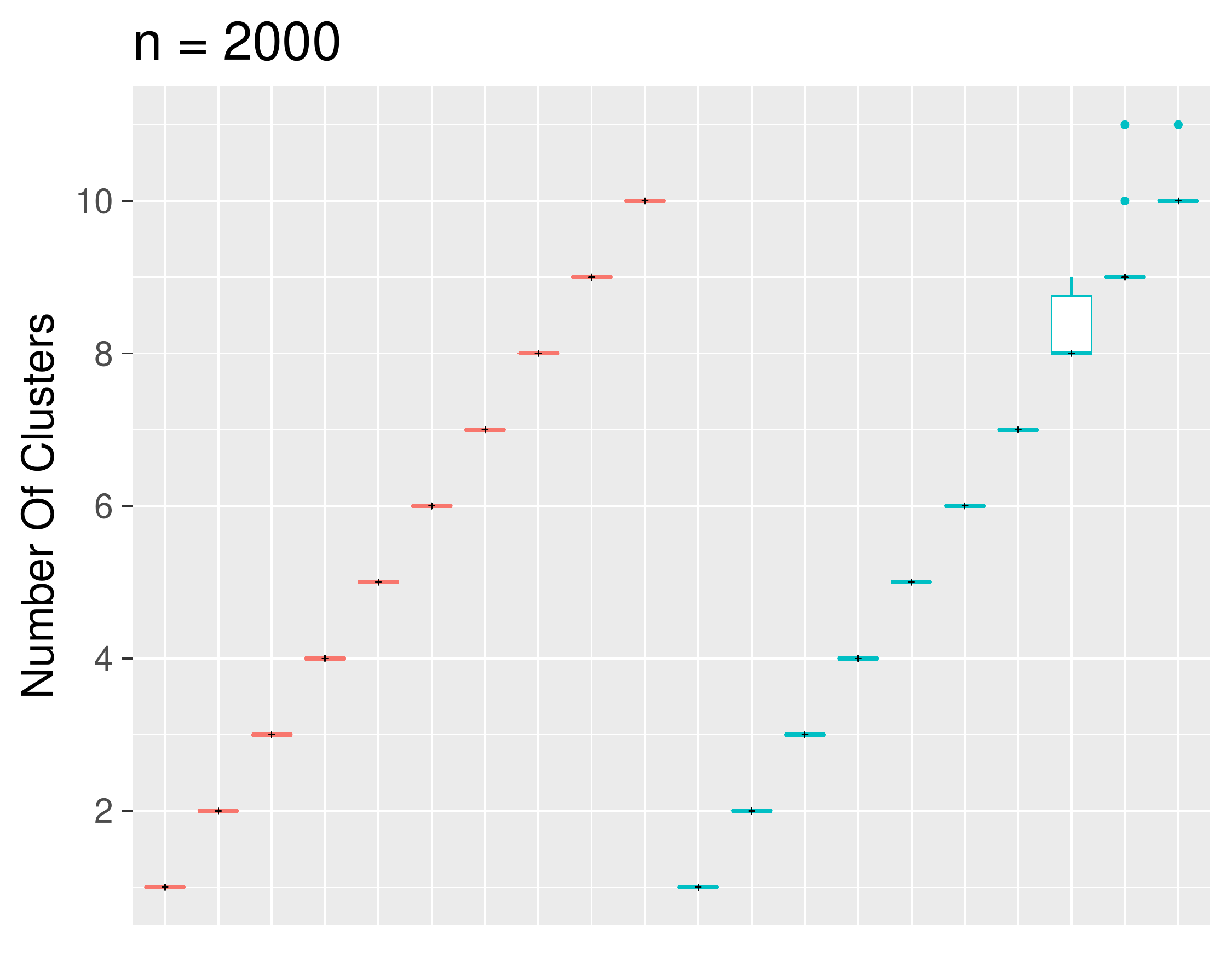} & \hspace{-3ex}\includegraphics[scale=0.3]{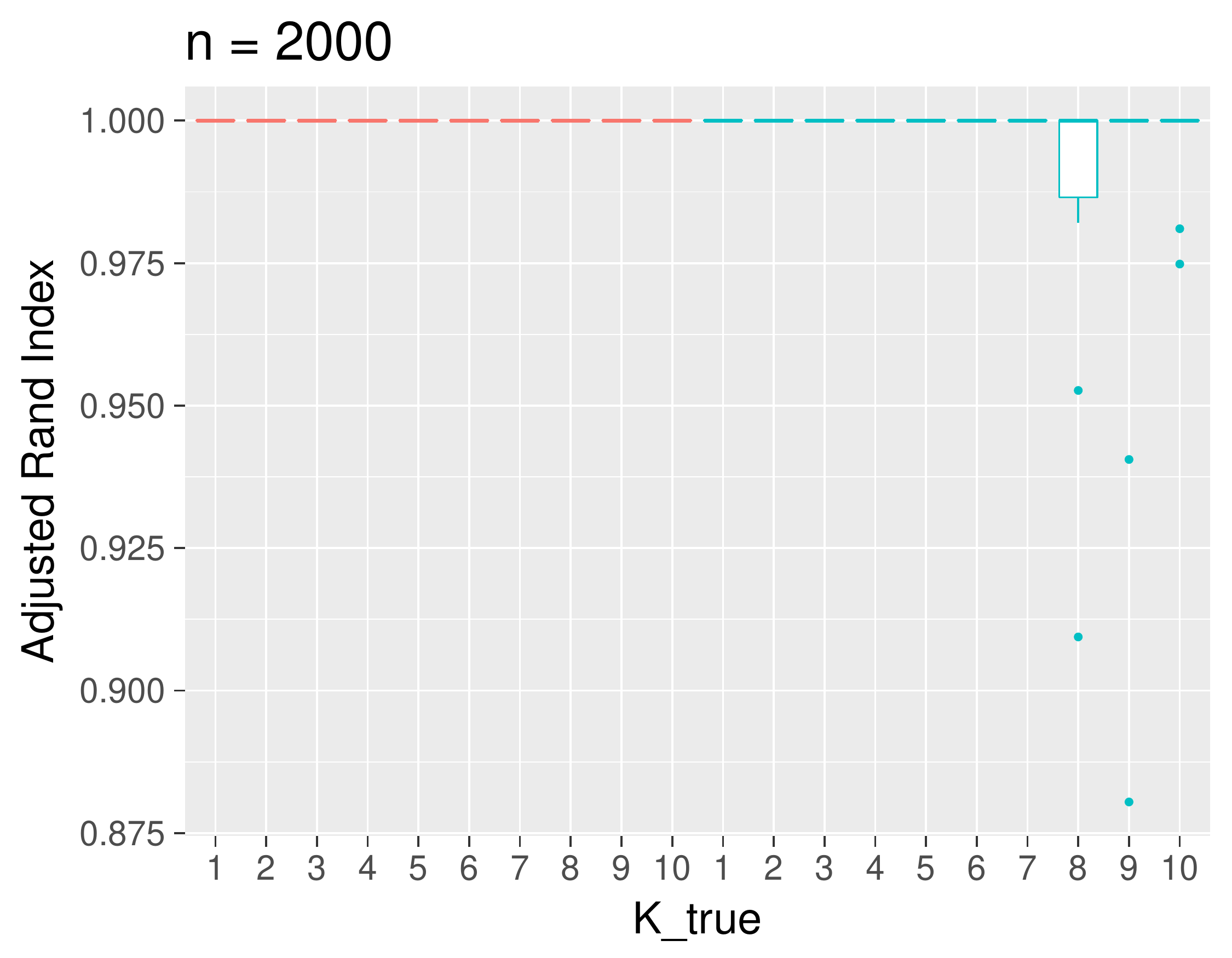}\end{tabular}\\
\hspace{-2ex}\includegraphics[scale=0.2]{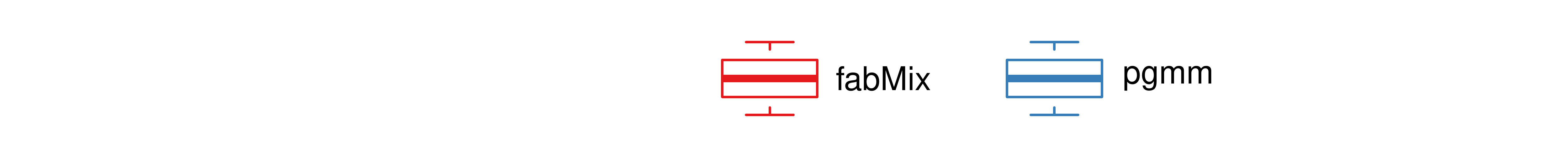} 
\caption{Left: Benchmarking MFA algorithms in terms of estimation of the number of clusters on synthetic data. We considered that the number of clusters ($K$) varies in the set $K\in\{1,\ldots,10\}$ and for each case ten different datasets were simulated. The cross denotes the real value of $K$. Right: Corresponding Adjusted Rand Index between the estimated single best clustering and the true cluster labels.
}
\label{fig:kSelection}
\end{figure}

Next, we repeat scenario 1 for a total of 300 synthetic datasets, considering that the true number of clusters ranges in the set $K_0\in\{1,\ldots,10\}$. For each value of $K_0$ we generated 10 datasets consisting of $n = 500$ observations, 10 datasets with $n = 1000$ and 10 datasets with $n = 2000$. As shown in Figure \ref{fig:kSelection} the proposed method is able to infer the true number of clusters as well as the correct clustering structure in terms of the ARI. Under six different starts, the EM algorithm implementation in {\tt pgmm} gives quite accurate results but note that sometimes the number of clusters may be overestimated when $K\geqslant 9$ and $n = 500, 1000$ (as shown in the first row of Table \ref{tab:overlapping}). Both methods always select the parameterization that corresponds to the one used to generate the data. More specifically, {\tt fabMix} selects the model with the same variance of errors per cluster and {\tt pgmm } selects the model corresponding to the {\tt "UCU"} parameterization (excluding a single case where the {\tt "UUU"} model was selected).


\begin{figure}[t]
\centering\includegraphics[scale=0.45]{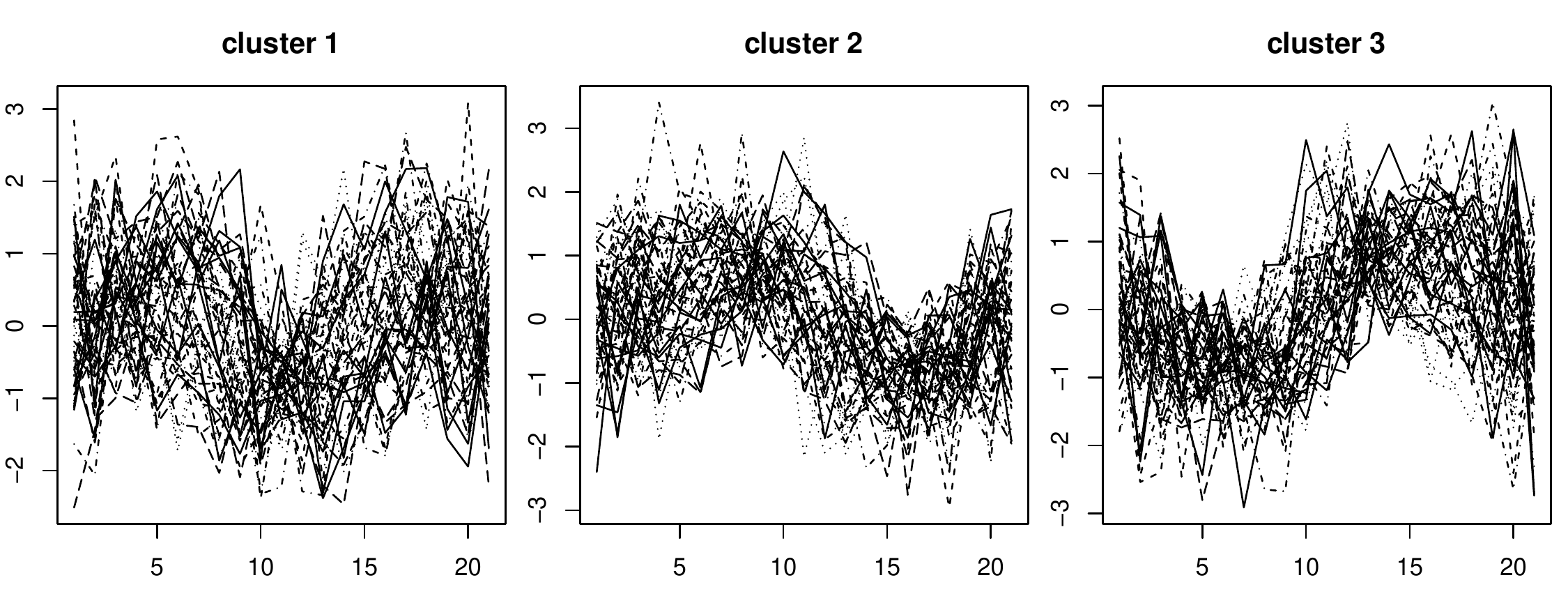}
\\
\centering\includegraphics[scale=0.45]{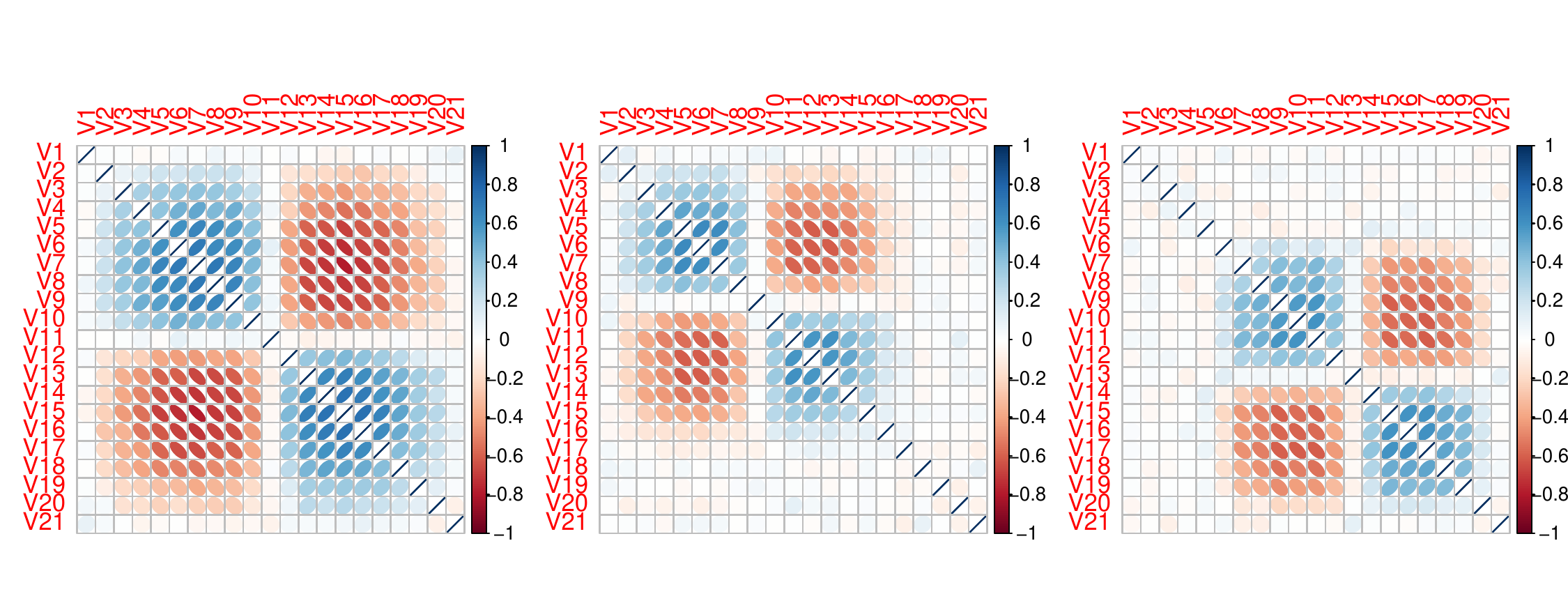}
\caption{Upper panel: Standardized wave dataset grouped according to the true cluster labels (showing only a randomly sampled subset of 50 observations per class). The raw data has been transformed prior to grouping so that the overall sample means and variances of the $p = 21$ variables are equal to zero and one, respectively. Lower panel: Sample correlation matrix per cluster.
}
\label{fig:wave_cluster}
\end{figure}

\subsection{Wave dataset}\label{sec:realData}

We selected a randomly sampled subset of 1500 observations from the wave dataset \citep{wavedata}, available from the UCI machine learning repository \citep{uci}. According to the available ``ground-truth'' classification of the dataset, there are 3 equally weighted underlying classes of $21$-dimensional continuous data, where each one is generated from a combination of 2 of 3 ``base'' waves, shown in Figure \ref{fig:wave_cluster}. The lower panel displays the sample correlation matrix per cluster, using the {\tt corrplot} package \citep{corrplot}.

\begin{table}[t]
\centering
\begin{tabular}{cllllll}
\toprule
& \multicolumn{3}{c}{{\tt fabMix}}
  & \multicolumn{3}{c}{{\tt pgmm}}
\\ 
 & 1& 2&  3 &  1 &  2 &  3\\
  \midrule
cluster 1 & 397 & 53 & 56 & 401 & 49 & 56\\ 
cluster 2 & 30 & 411 & 40& 31& 409 & 41\\ 
cluster 3 & 15 &  22 & 476& 17 & 22 & 474\\   \midrule
\# factors (BIC) & \multicolumn{3}{c}{1} &  \multicolumn{3}{c}{1} \\ \midrule
ARI & \multicolumn{3}{c}{0.616} &  \multicolumn{3}{c}{0.615} \\ \midrule
RI & \multicolumn{3}{c}{0.829} &  \multicolumn{3}{c}{0.829} \\ \bottomrule
\end{tabular}
\caption{Confusion matrix between the estimated and true cluster assignments of the wave dataset. All methods select $1$ factor and $3$ clusters. The last two lines  contain the adjusted and raw Rand index of each method with respect to the true cluster labels.}
\label{table1}
\end{table}

\begin{figure}[!ht]
\centering\includegraphics[scale=0.45]{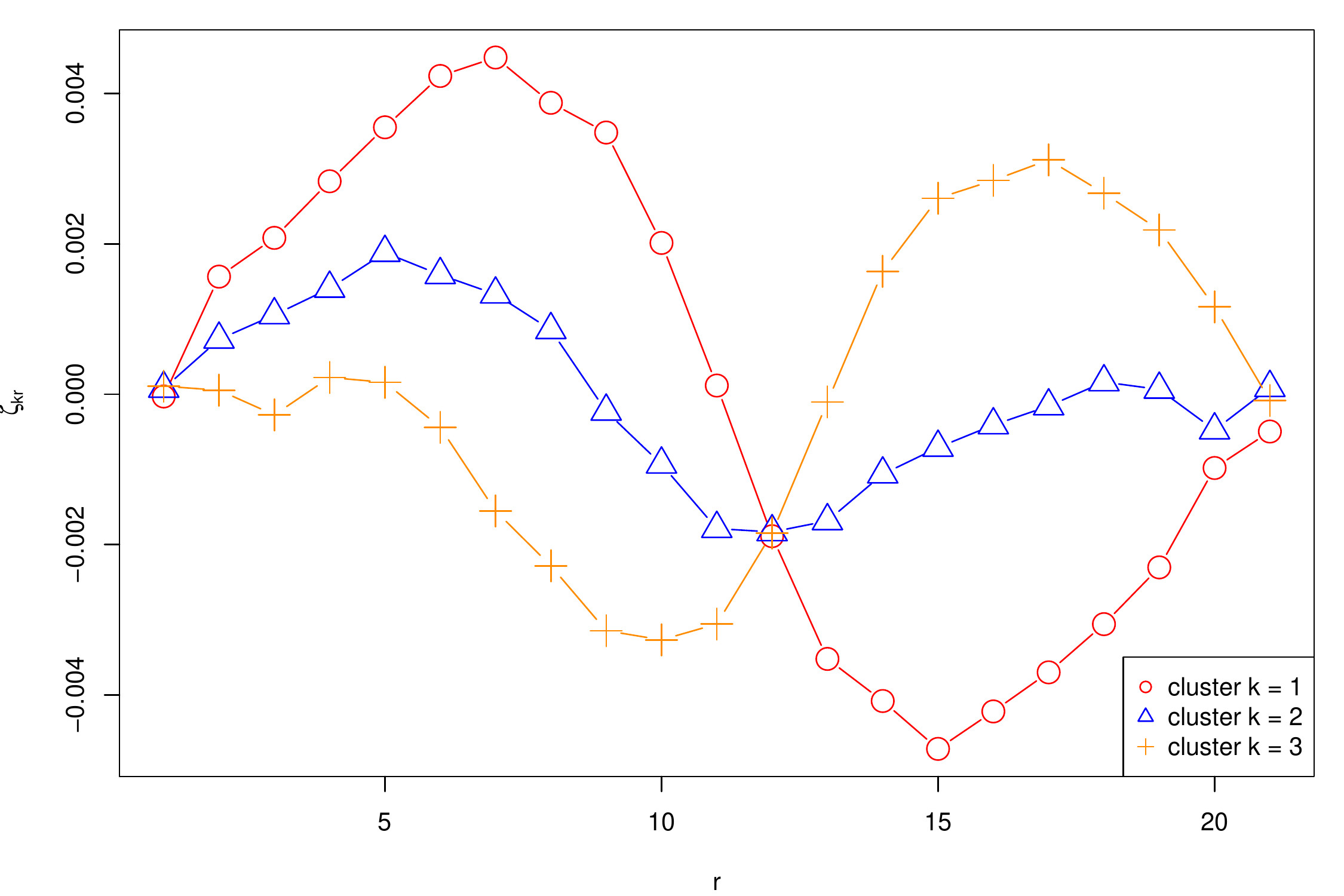}
\caption{Interpretation of factor effects per cluster for the wave dataset. The selected model corresponds to $q = 1$ factor and $K = 3$ clusters. The lines represent the estimated ``regularised score'' $\widehat{\zeta}_{kr}$ (see Equation \eqref{eq:regExp}) of each variable $r = 1,\ldots,p$ to the (single) factor for cluster $k = 1,2,3$. 
}
\label{fig:zetas}
\end{figure}

We applied the proposed method considering that the number of factors ranges in the set $q\in\{1,\ldots,4\}$. According to BIC, the selected number of factors is equal to $q = 1$. Conditionally on $q = 1$, our method selects the true value of clusters as the most probable number of ``alive components''. In particular the corresponding  estimated posterior probability is $\widehat{\rm{P}}(K_0 = 3|\bs x) = 0.64$. The same model is also chosen by {\tt pgmm}. The estimated clusters with respect to the true allocation of each observation is shown in Table \ref{table1}. It is evident that the resulting estimates are very similar to each other (notice that {\tt pgmm} and {\tt fabMix} achieve the same ARI) and exhibit strong agreement with the true clustering of the data.  {\tt fabMix} selects the parameterization with the same variance of errors per component and {\tt pgmm} selects the {\tt "UCU"} parameterization. 


As far as factor interpretation is concerned, we estimated the ``regularized score'' of each variable to the (single) factor per cluster. For this purpose, we only considered the simulated values that correspond to three alive components and reorder the MCMC values to undo the label switching. Then, we take ergodic averages in order to estimate $\zeta_{krj} = \zeta_{kr}$ (single factor $j = 1$) in \eqref{eq:regExp} for $k = 1,2,3$ and $r = 1,\ldots,21$, as shown in Figure \ref{fig:zetas}. For the first cluster ($k=1$) we note that $\widehat{\zeta}_{1r}$ takes positive values for $2\leqslant r\leqslant 10 $ and negative values for $12\leqslant r\leqslant 20 $, while $\widehat{\zeta}_{1r} \approx 0$ for $r = 1, 11, 21$. Note the strong agreement when comparing these values with the patterns in the correlation matrix in Figure \ref{fig:wave_cluster} for true cluster 1. It is evident that the block of variables $\{2,\ldots,10\}$ has different correlation behaviour than  $\{12,\ldots,20\}$. Moreover, observe that the block of variables $\{1, 11, 21\}$ is not correlated to any of the other variables. Similar conclusions can be drawn when comparing the regularized scores for clusters 2 and 3 with their correlation structure.

\subsection{Italian wines}\label{sec:wines}

The wine dataset \citep{forina1986multivariate}, available at the {\tt pgmm} package \citep{pgmm}, contains $p=27$ variables measuring chemical and physical properties of $n = 178$ wines. There are three types of wine (Barolo, Grignolino, Barbera) from the Piedmont region of Italy which were collected over the ten year period 1970--1979. We applied our method considering $K = 20$ and $1\leqslant q \leqslant 5$. For the same number of factors we also applied {\tt pgmm}, considering that the number of components ranges in the set $1\leqslant K\leqslant 5$. The number of initializations and heated chains are the same as in Section \ref{sec:realData}. As shown in Table \ref{tab:wine}, {\tt fabMix} selects a model with 4 clusters and 2 factors and {\tt pgmm} selects a model with 3 clusters and 4 factors. The best clustering performance corresponds to {\tt pgmm}, which achieves an almost perfect classification. Furthermore, we report that {\tt fabMix} selects the parameterization with different variance of errors per component, while {\tt pgmm} selects the {\tt "CUU"} parameterization. 


\begin{table}[t]
\centering
\begin{tabular}{clllllll}
\toprule
& \multicolumn{4}{c}{{\tt fabMix}}
  & \multicolumn{3}{c}{{\tt pgmm}}
\\ 
 & 1& 2& 3 & 4& 1 &  2 & 3  \\
  \midrule
cluster 1 (Barolo)     & 0 & 48 & 11 & 0  & 0 & 59 & 0 \\ 
cluster 2 (Grignolino) & 1 & 18 & 3  & 49 & 1 & 1 &69  \\ 
cluster 3 (Barbera)    & 48  &0 & 0  & 0  & 48 & 0 &0 \\ \midrule
\# factors (BIC) & \multicolumn{4}{c}{2} &  \multicolumn{3}{c}{4}  \\ \midrule
ARI & \multicolumn{4}{c}{0.614} &  \multicolumn{3}{c}{0.965}  \\ \midrule
RI & \multicolumn{4}{c}{0.830} &  \multicolumn{3}{c}{0.984}  \\ \bottomrule
\end{tabular}
\caption{Confusion matrix between the estimated and true cluster assignments of the wine dataset. The last two lines  contain the adjusted and raw Rand index of each method with respect to the true cluster labels.}
\label{tab:wine}
\end{table}

\subsection{Coffee dataset}\label{sec:coffee}

The coffee dataset \citep{Streuli} consists of $n=43$ coffee samples collected from beans corresponding to the Arabica and Robusta species. For each sample thirteen variables are observed: water, pH value, fat, chlorogenic acid, bean weight, free acid, caffeine, neochlorogenic acid, extract yield, mineral content, trigonelline, isochlorogenic acid and total chlorogenic acid. The dataset has been previously analyzed using mixtures of factor analyzers by \cite{McNicholas2008} and is available at the {\tt pgmm} package \citep{pgmm}. Following \cite{McNicholas2008}, the total chlorogenic acid is excluded from the analysis since it is the sum
of the chlorogenic, neochlorogenic and isochlorogenic acid values. 

\begin{table}[t]
\centering
\begin{tabular}{cllllllll}
\toprule
& \multicolumn{2}{c}{{\tt fabMix}}
  & \multicolumn{5}{c}{{\tt pgmm}}
\\ 
 & 1& 2 & 1 &  2 & 3 & 4& 5\\
  \hline
cluster 1 (Arabica) & 36 & 0 &0 &6& 4 & 10& 16  \\ 
cluster 2 (Robusta) & 0  & 7 &7 & 0 & 0 & 0 & 0 \\ \midrule
\# factors (BIC) & \multicolumn{2}{c}{1} &  \multicolumn{5}{c}{3} \\ \midrule
ARI & \multicolumn{2}{c}{1} &  \multicolumn{5}{c}{0.206}  \\ \midrule
RI & \multicolumn{2}{c}{1} &  \multicolumn{5}{c}{0.508}  \\ \bottomrule
\end{tabular}
\caption{Confusion matrix between the estimated and true cluster assignments of the coffee dataset. The last two lines  contain the adjusted and raw Rand index of each method with respect to the true cluster labels.}
\label{tab:coffee}
\end{table}

We applied our method considering $K = 20$ and $1\leqslant q \leqslant 5$. For the same number of factors we also applied {\tt pgmm}, considering that the number of components ranges in the set $1\leqslant K\leqslant 5$. As shown in Table \ref{tab:coffee}, {\tt fabMix} selects a model with 2 clusters and 1 factor and {\tt pgmm} selects a model with 5 clusters and 3 factors. The best clustering performance in terms of the adjusted and raw Rand indexes corresponds to {\tt fabMix} which achieves a perfect classification. Furthermore, we report that {\tt fabMix} selects the parameterization with different variance of errors per component and {\tt pgmm} selects the {\tt "CCUU"} parameterization. 

We observed that in this case the estimates obtained by the EM algorithm are quite sensitive to the selection of starting values, a behaviour which typically indicates a highly multimodal likelihood surface. Indeed, based on 100 independent calls to {\tt pgmm} consisting of six initializations under different values, the algorithm selected a model with $K = 2,3,4,5$ clusters with frequencies equal to $41, 1, 30, 28$, respectively. For this reason, the results reported in Table \ref{tab:coffee} are based on a very large number (1000) of random starting values for {\tt pgmm} (as well as one run based on $K$-means starting values). We note that a perfect classification was achieved when {\tt pgmm} selected a model with $K = 2$ clusters, as reported in Table 15 of \cite{McNicholas2008}. For each of the 41 runs which selected a model with $K = 2$, the {\tt "CCUU"} parameterization was returned with a BIC value equal to  $1292.821$. But the {\tt pgmm} results in Table \ref{tab:coffee} correspond to a smaller BIC value ($840.27$) (thus, a better model). Finally, we mention that the same model is still ranked first when using the ICL criterion \citep{Biernacki2000}.

\section{Discussion and further remarks}\label{sec:discussion}

This study presented a solution to the complex problem of clustering multivariate data using Bayesian mixtures of factor analyzers. The proposed model builds upon the prior assumptions of \cite{Fokoue2003} assuming a fixed number of clusters and factors. Extending the Bayesian framework of \cite{overfitting} we demonstrated that estimating an overfitting mixture model is a straightforward and efficient approach to the problem at hand. We also used prior parallel tempering schemes in order to improve the mixing of the algorithm. The posterior inference conditionally on a specific model (number of alive clusters) is possible after applying suitable algorithms to deal with label switching.  

According to our simulation study, the proposed method can accurately infer the number and composition of the underlying clusters. Our results were benchmarked against state-of-the-art software for estimating MFA models via the EM algorithm, that is, the R package {\tt pgmm}. Based on six starts of {\tt pgmm}, we concluded that our method is quite competitive with the EM algorithm in estimation accuracy and can lead to improved inference as the number of clusters grows large. Starting values and the usage of multiple runs are very important for maximum likelihood methods. At this point recall that our benchmarking simulation study generated 100 datasets with dimension $500\times 40$, 100 datasets with dimension $1000\times 40$,  and 100 datasets with dimension $2000\times 40$. Using a single EM run for each dataset, the number of compared models for a given parameterization is equal to $K_{\mbox{max}}\times Q_{\mbox{max}}$, with $K_{\mbox{max}} = 20$ (maximum number of components) and $Q_{\mbox{max}}$ denoting the maximum number of factors, hence this number should be multiplied by the total number of runs in the case of initializing the EM from multiple starting values (as done in {\tt pgmm}). On the contrary, the total number of compared models for each parameterization under our Bayesian approach is equal to $Q_{\mbox{max}}$.

We are currently extending the method to the case where the multivariate data contains missing values. This is straightforward to implement under our Bayesian set-up by adding one extra Gibbs sampling step that  generates samples from the full conditional distribution in Equation \eqref{eq:fullConditionalX}. Thus, our future work will focus on presenting, benchmarking and applying  this extra functionality to real datasets, as well as improving the practical implementation with the addition of user-friendly output summaries and plot methods. Another interesting generalization of our model would be to propose Bayesian estimation of overfitted mixture models using the whole set of  parameterizations introduced in the family of EPGMMs. In this case the dimensionality of the parameter space is reduced further, thus, it is expected that our method will provide more flexible results in certain applications (such as the Italian Wine dataset where {\tt pgmm} performed better than {\tt fabMix}).

The source code of the proposed algorithm is hosted online at \url{https://github.com/mqbssppe/overfittingFABMix} in the form of an {\tt R} package, together with scripts that reproduce the results of Section \ref{sec:results}. Future versions of the package will be also available on CRAN. 

\section{Supplementary Material}

\ref{S1_Appendix} contains some technical details regarding the implementation of the Gibbs sampler. A detailed description of the generation of synthetic data can be found in \ref{S2_Appendix}. \ref{sec:standardBMFA} compares the proposed overfitting Bayesian MFA model against standard Bayesian MFA models. \ref{sec:convergence} discusses convergence diagnostics and comparison with a simpler initialization scheme. Detailed results on the estimation of the number of factors is given in \ref{sec:qEstimation}.

\section{Acknowledgments}
Research was funded by MRC award MR/M02010X/1. The author would like to acknowledge the assistance given by IT services and use of the Computational Shared Facility of the University of Manchester. The suggestions of two anonymous reviewers helped to improve the findings of this study.

\appendix
\section{Supplementary Material}

\subsection{Gibbs sampling updates}\label{S1_Appendix}  
In the following, $(x|y)$ denotes the distribution of  $x$ conditionally on $y$. Given $K$ and $q$ the Gibbs sampler updates each parameter according to the following scheme.

\begin{enumerate}
\item Give some initial values $(\bs\Omega^{(0)}, \bs\Lambda^{(0)}, \bs\mu^{(0)},  \bs z^{(0)}, \bs\Sigma^{(0)}, \bs  w^{(0)},  \bs y^{(0)})$.
\item At iteration $t = 1,2,\ldots$
\begin{enumerate}
\item update $\bs\Omega^{(t)}\sim\left(\bs\Omega|\bs\Lambda^{(t-1)}\right)$
\item update $\bs\Lambda^{(t)}\sim\left(\bs\Lambda|\bs\Omega^{(t)}, \bs\Sigma^{(t-1)}, \bs x, \bs y^{(t-1)}, \bs z^{(t-1)}\right)$
\item update $\bs\mu^{(t)}\sim\left(\bs\mu|\bs\Lambda^{(t)},\bs \Sigma^{(t-1)}, \bs x, \bs y^{(t-1)}, \bs z^{(t-1)}\right)$
\item update $\bs z^{(t)}\sim \left(\bs z|\bs w^{(t-1)},\bs \mu^{(t)}, \bs\Lambda^{(t)}, \bs\Sigma^{(t-1)}, \bs x\right)$
\item update $\bs\Sigma^{(t)}\sim\left(\bs\Sigma|\bs x, \bs z^{(t)}, \bs\mu^{(t)}\right)$
\item update $\bs w^{(t)}\sim\left(\bs w|\bs z^{(t)}\right)$
\item update $\bs y^{(t)}\sim\left(\bs y|\bs x, \bs z^{(t)}, \bs\mu^{(t)}, \bs\Sigma^{(t)},\bs \Lambda^{(t)}\right)$.
\end{enumerate}
\end{enumerate}
All full conditional distributions are detailed in \cite{Fokoue2003}. However, in some cases, it is easy to analytically integrate some parameters and accelerate  convergence. Note that in steps (2.b) and (2.d) we have integrated out $\bs\mu$ and $ \bs y$, respectively. Thus, in steps (2.b) and (2.d), the parameters are updated from the corresponding collapsed conditional distributions, rather than the full conditional distributions. 

In particular, after integrating out $\bs\mu = (\bs\mu_1,\ldots,\bs\mu_K)$, it follows that
\begin{equation}\label{eq:lambda_update}
(\bs\Lambda_{kr\cdot}|\bs\Omega, \bs\Sigma, \bs x,\bs y,\bs z) = \mathcal N_q(\bs\Gamma_{kr}\bs\Delta_{kr}^T, \bs\Gamma_{kr}),
\end{equation}
independent for $k = 1,\ldots,K$; $r = 1,\ldots,p$. The parameters on the right-hand side of Equation \eqref{eq:lambda_update} are defined as
\begin{eqnarray*}
\bs\Gamma_{kr} &=& \left(\bs\Omega_{\nu_r}^{-1} + \frac{1}{\sigma^{2}_{kr}}\ddot{s}_k(\bs z,\bs  y) - \frac{\phi_{krr}}{\sigma^{4}_{kr}}\ddot{s}_{k;1:\nu_r}(\bs z, \bs y)\ddot{s}_{k;1:\nu_r}^{T}(\bs z,\bs y)\right)^{-1}\\
\bs\Delta_{kr} &=& \frac{1}{\sigma_{kr}^{2}}\dddot{s}_{kr}(\bs z, \bs x,\bs  y) - \frac{\phi_{krr}}{\sigma^{4}_{kr}}\ddot{s}_{kr}(\bs z,\bs  x)\ddot{s}_{k;1:\nu_r}^{T}( \bs z, \bs y)
\end{eqnarray*}
where
\begin{eqnarray*}
\dot{s}_k(\bs z) &=& \sum_{i=1}^{n}I(z_i = k)\\
\ddot{s}_k(\bs z,\bs u) &=& \sum_{i=1}^{n}I(z_i = k)\bs u_i \bs u_i^{T}\\
\dddot{s}_k(\bs z,\bs  u, \bs v) &=& \sum_{i=1}^{n}I(z_i = k)\bs u_i \bs v_i^{T}\\
\bs \Phi_{k} &=& \left(\dot{s}_k(\bs z)\bs\Sigma_k^{-1} +\bs\Psi^{-1}\right)^{-1},
\end{eqnarray*}
for $k = 1,\ldots,K$. Finally, from Equations \eqref{eq:z_prior} and \eqref{eq:mixture} it immediately follows that 
\begin{equation*}
\mathrm{P}(z_i = k|\bs x_i, \bs\mu, \bs\Sigma, \bs\Lambda,\bs w) \propto w_kf\left(\bs x_i;\bs\mu_k,\bs\Lambda_k\bs\Lambda_k^{T} + \bs\Sigma_k\right), \quad k=1,\ldots,K,
\end{equation*}
independent for $i  =1,\ldots,n$, where $f(\cdot;\bs\mu,\bs\Sigma)$ denotes the probability density function of the multivariate normal distribution with mean $\bs\mu$ and covariance matrix $\bs\Sigma$. Note that in order to compute the right hand side of the last equation, inversion of the $p\times p$ matrix $\bs\Lambda_k\bs\Lambda_k^{T} + \bs\Sigma_k$ is required. Since $\bs\Sigma_k$ is diagonal, this matrix can be efficiently inverted using the Sherman--Morrison--Woodbury formula \citep{10.2307/2030425}:
$$\left(\bs\Lambda_k\bs\Lambda_k^{T} + \bs\Sigma_k\right)^{-1} = \bs\Sigma_k^{-1} - \bs\Sigma_k^{-1}\bs\Lambda_k\left(\bs{\mathrm{I}}_q + \bs\Lambda_k^T\bs\Sigma^{-1}_k\bs\Lambda_k\right)^{-1}\bs\Lambda_k^T\bs\Sigma^{-1}_k,$$
for $k = 1,\ldots,K$. 

\subsection{Simulation study details}\label{S2_Appendix} 
The synthetic data in Section \ref{sec:simulationStudy} was simulated according to mixtures of multivariate normal distributions, as shown in Equation \eqref{eq:mixture}. Given the number of mixture components ($K$) and number of factors ($q$) for each dataset, the true values of parameters were generated according to the following scheme. 
\begin{eqnarray*}
\bs w&\sim&\mathcal D(10,\ldots,10)\\
\mu_{kr} &=& \begin{cases}
20\sin\{\frac{r-1}{p-1}k\pi\},& \mbox{ with probability } 1/3,\\
20\cos\{\frac{r-1}{p-1}k\pi\},& \mbox{ with probability } 1/3,\\
-40\cos\{\frac{r-1}{p-1}2k\pi\},& \mbox{ with probability } 1/3,
\end{cases}\\
\sigma_{krr}^{2} &=& \begin{cases}r, & \mbox{Scenario 1} \\100, & \mbox{Scenario 2} \\ 0.1, & \mbox{Scenario 3} \end{cases}\\
\Lambda_{k}^{T} &=&
\left[
\begin{array}{cccc|cccc|c|cccc|ccc}
\bigstar & \bigstar & \cdots & \bigstar & 
\square & \square & \cdots & \square & 
\cdots &
\square & \square & \cdots & \square &
\square & \cdots &\square
\\

\square & \square & \cdots & \square & 
\bigstar & \bigstar & \cdots & \bigstar &  
\cdots &
\square & \square & \cdots & \square &
\square & \cdots &\square\\

\vdots & \vdots & \ddots & \vdots &
\vdots & \vdots & \ddots & \vdots &
\ddots &
\vdots & \vdots & \ddots & \vdots &
\vdots & \ddots & \vdots \\

\square & \square & \cdots & \square & 
\square & \square & \cdots & \square & 
\cdots &
\bigstar & \bigstar & \cdots & \bigstar &
\square & \cdots &\square\\

\end{array}
\right],
\end{eqnarray*}
independent for $r=1,\ldots,p$; $k = 1,\ldots,K$. Observe that the diagonal matrix containing the variance of errors ($\sigma^2_{krr}$) is the same for each  mixture component. Under a slight abuse of notation in the $q \times p$ matrix $\bs\Lambda_k^T$, a star ($\bigstar$) denotes independent realizations from a random variable following a 
\begin{equation}\label{eq:loadingsDist}
\mathcal N(\phi_k, \rho^2_k)
\end{equation}
distribution. Moreover, a square ($\square$) denotes independent realizations from a $\mathcal N(0, 1)$ random variable. Each block matrix containing a star consists of $q$ rows and $\lfloor p/q\rfloor$ columns, where $\lfloor \cdot\rfloor$ denotes the integer part of a positive number. In case that $p/q$ is not integer, we fill the remaining $p - q\lfloor p/q \rfloor$ columns with squares (corresponding to the last block of $\bs\Lambda_k^{T}$). The parameters of the distribution  \eqref{eq:loadingsDist} are generated as follows:
\begin{equation*}
\phi_k \sim \begin{cases}
\mathcal{DU}\{-30,-20,-10,10,20,30\}, & \mbox{Scenario 1}\\
\mathcal{DU}\{-30,-20,-10,10,20,30, 1, 40\},& \mbox{Scenario 2}\\
1,& \mbox{Scenario 3}
\end{cases}
\end{equation*}
and
\begin{equation*}
\rho_k = \begin{cases}
2, & \mbox{Scenario 1}\\
0.2|\phi_k| + 1,& \mbox{Scenario 2}\\
0.01,& \mbox{Scenario 3}
\end{cases}
\end{equation*}
for $k =1,\ldots,K$. In all cases $\phi_k$ and $\rho_k$  are assumed independent for $k=1,\ldots,K$.

\begin{figure}[t]
\centering
\begin{tabular}{cc}
\includegraphics[scale=0.39]{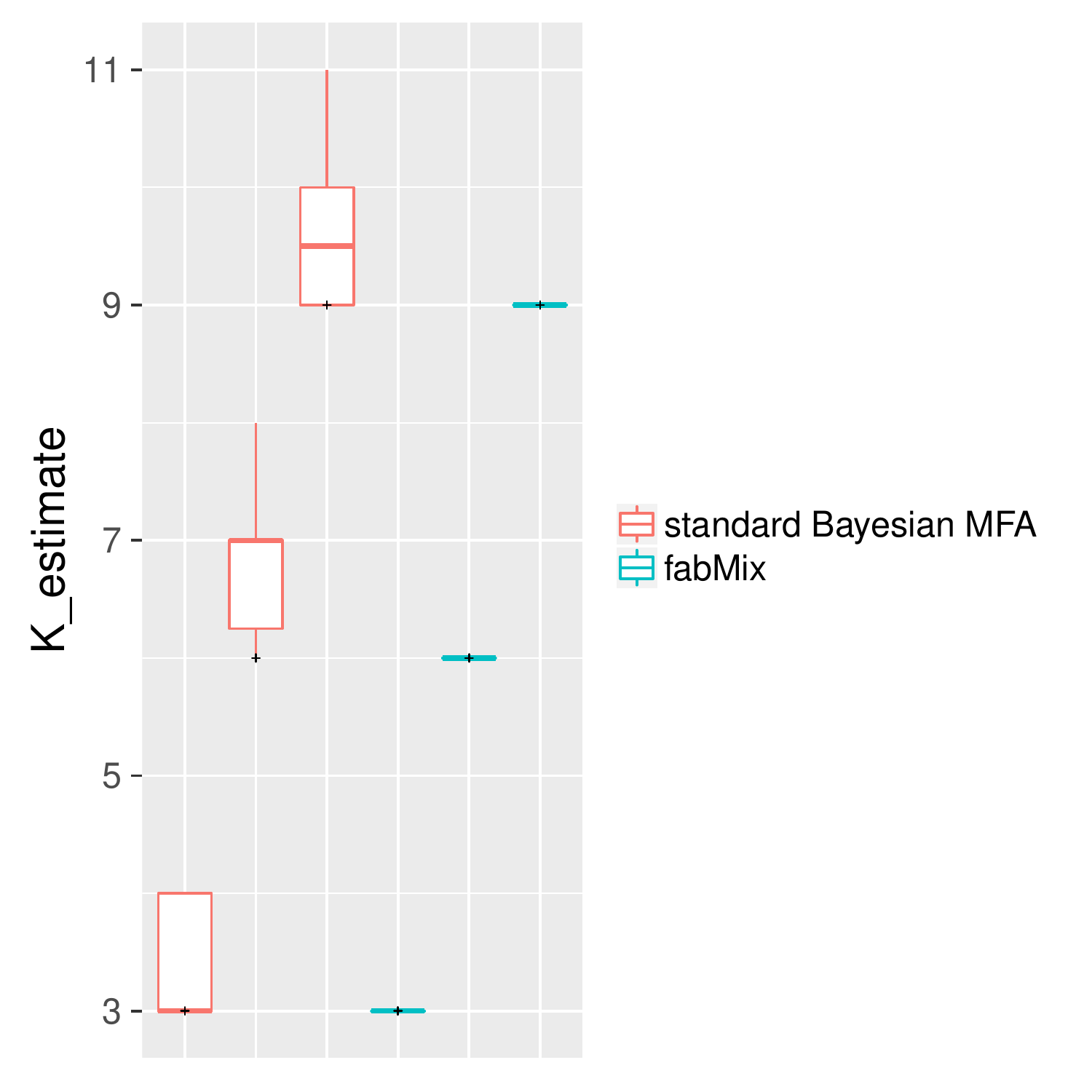} &
\includegraphics[scale=0.39]{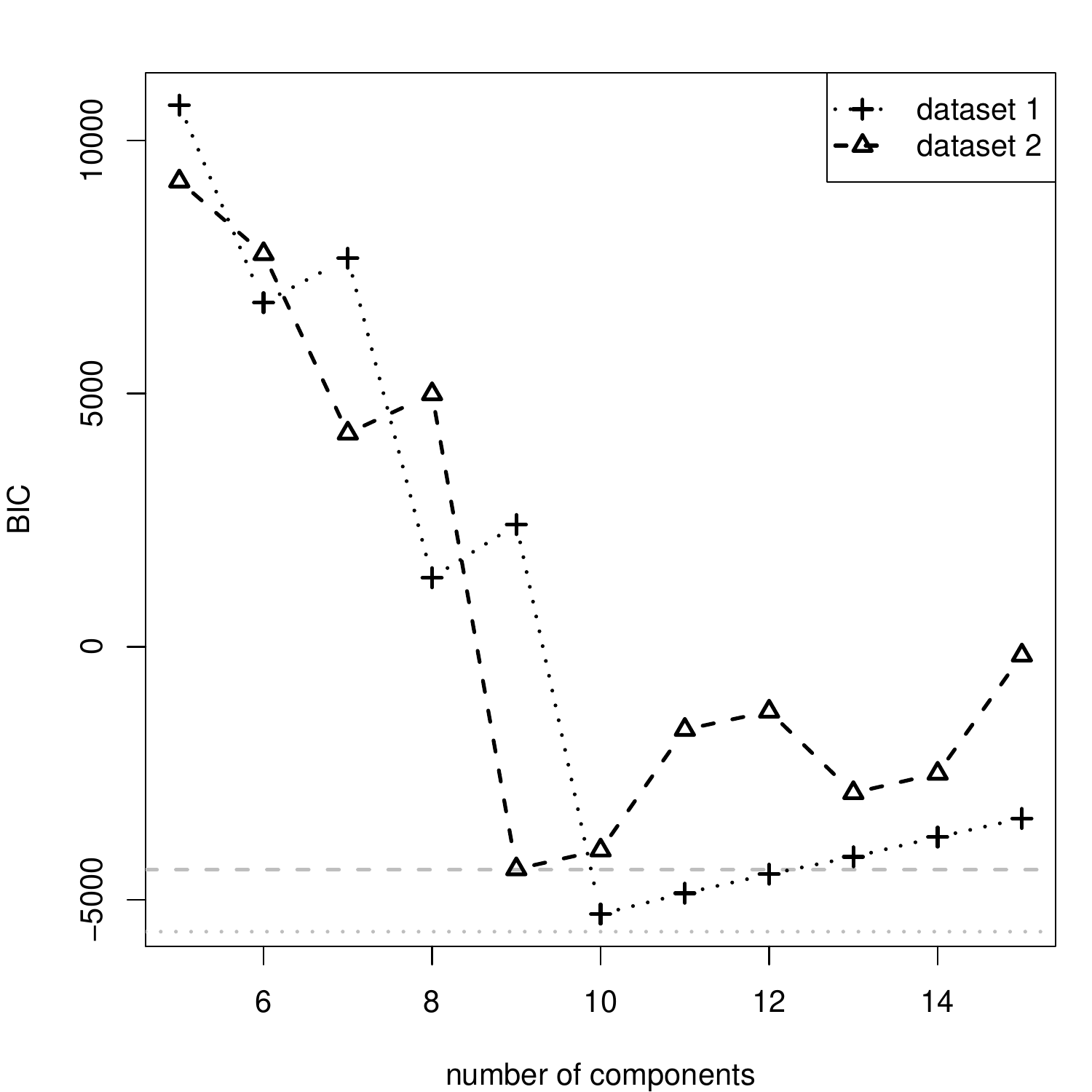}\\
(a) & (b)
\end{tabular}
\caption{(a): Benchmarking overfitting against standard Bayesian MFA models in terms of estimation of the number of clusters on synthetic data. We considered that the number of clusters ($K$) varies in the set $K\in\{3,6,9\}$ and for each case ten different datasets were simulated. The cross denotes the real value of $K$. (b): BIC values obtained by the standard Bayesian MFA model for two simulated datasets with true number of components equal to $K = 9$. The horizontal gray lines indicate the BIC value obtained by the corresponding overfitting Bayesian MFA model, conditionally on the values of the most probable number of ``alive'' components. 
}
\label{fig:standardBMFA}
\end{figure}

\subsection{Benchmarking against standard initialization and Bayesian MFA models}\label{sec:standardBMFA}

In this section we compare against standard Bayesian MFA models. We considered $30$ simulated datasets with $n = 500$ and $p = 20$ with a fixed number of factors $q = 2$, which is assumed known. Then, we estimated the number of clusters using the BIC by fitting a series of standard Bayesian MFA models with a number of components equal to $K=1,\ldots,20$. The results are compared against the proposed method using $J = 8$ parallel chains. Note that no parallel chains are considered in the standard Bayesian MFA MCMC sampler. In order to make the comparison as fair as possible, we considered that the number of MCMC iterations is equal to $Jm$ and $m$ for standard and overfitting MFA, respectively, with $m = 20000$ as in the previous sections. 

As shown in Figure \ref{fig:standardBMFA}(a), the proposed method gives more accurate results than standard Bayesian MFA models, especially as the number of components grows large. Figure \ref{fig:standardBMFA}(b) displays the behaviour of the BIC values of the standard Bayesian MFA model for two specific datasets where the true number of components is equal to $K = 9$. Observe that the standard MCMC sampler selects $K=10$ clusters for dataset 1 and $K = 9$ clusters for dataset 2. Furthermore, note that the BIC values, as a function of the number of components, are not smooth for both datasets. Such a behaviour is typical in cases where the algorithm has converged to minor modes (see \cite{papastamoulis2016estimation}).

\subsection{Convergence diagnostics}\label{sec:convergence}

\begin{figure}[t]
\centering
\includegraphics[scale=0.39]{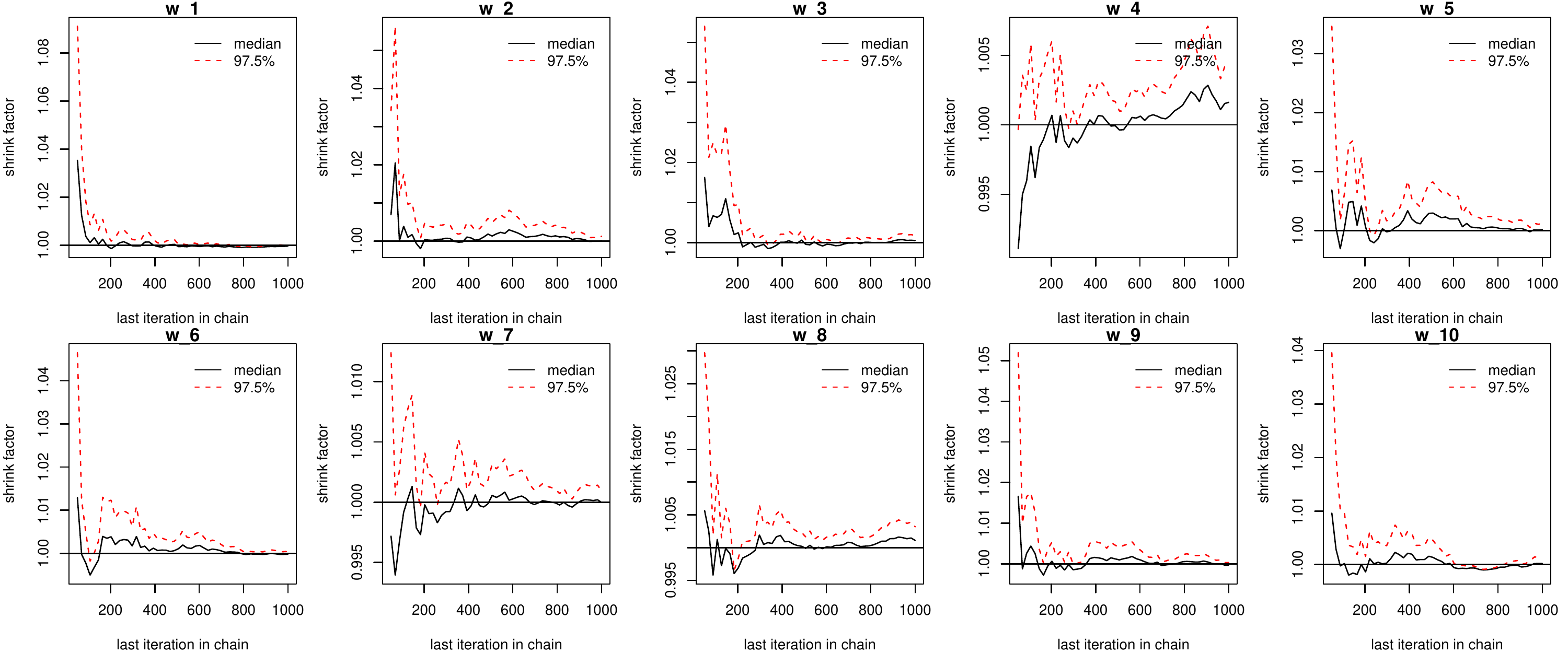} 
\caption{Convergence of mixing proportions $w_1,\ldots,w_{10}$: evolution of Gelman and Rubin's shrink factor as the number of iterations increases.  
}
\label{fig:gelmanplot}
\end{figure}

In order to evaluate the convergence of our algorithm, we have ran the sampler for the same number of iterations ($m=20000$) based on different starting values for a total of 10 runs using a simulated dataset with $K = 10$ clusters and $q = 4$ factors. The number of observations is equal to $n = 500$ for a total of $p = 40$ variables generated according to scenario 1. Using the {\tt coda} package \citep{coda}, the Gelman-Rubin diagnostic criterion \citep{gelman1992inference} varies between $(1.00, 1.012)$ (mixing proportions), $(1.00, 1.04)$ (means) and $(1.00, 1.13)$ (covariance matrices). Note that the clusters have been relabelled across all runs so that they agree to each other. This range of values is quite smaller than $1.2$, so according to \cite{brooks1998general} we can be fairly confident that convergence has been reached. Figure \ref{fig:gelmanplot} illustrates the evolution of the Gelman-Rubin diagnostic criterion for the mixing proportions as the number of iterations increases.

Furthermore, we compare the performance of our MCMC sampler when a different initialization scheme is applied. Recall that the default initialization scheme is based on a first stage of 100 iterations where the mixing proportions are a-priori distributed according to a Dirichlet distribution with parameters that lead to overfitting, as discussed in Section \ref{sec:details}. Now we consider a simpler scheme that does not include this stage and all parameters are randomly generated from the default prior distributions. Under the same number of iterations and independent runs, we calculated that the  Gelman-Rubin diagnostic criterion \citep{gelman1992inference} varies between $(1.00, 1.45)$ (mixing proportions), $(1.00, 1.86)$ (means) and $(1.15, 2.31)$ (covariance matrix). This means that the MCMC sampler has not converged when using this simpler initialization scheme, under the same number of iterations.

\subsection{Estimating the number of factors}\label{sec:qEstimation}

In this section we assess the ability of our algorithm to infer the number of factors. We generated synthetic data from mixtures of factor analyzers where the true number of factors varies in the set $q\in\{1,\ldots,5\}$ and for each value 15 datasets were simulated. The true number of clusters is uniformly distributed on the set $\{1,\ldots,10\}$. For each case we considered that $n = 1000$ and $p = 40$.

\begin{figure}[t]
\centering\includegraphics[scale=0.4]{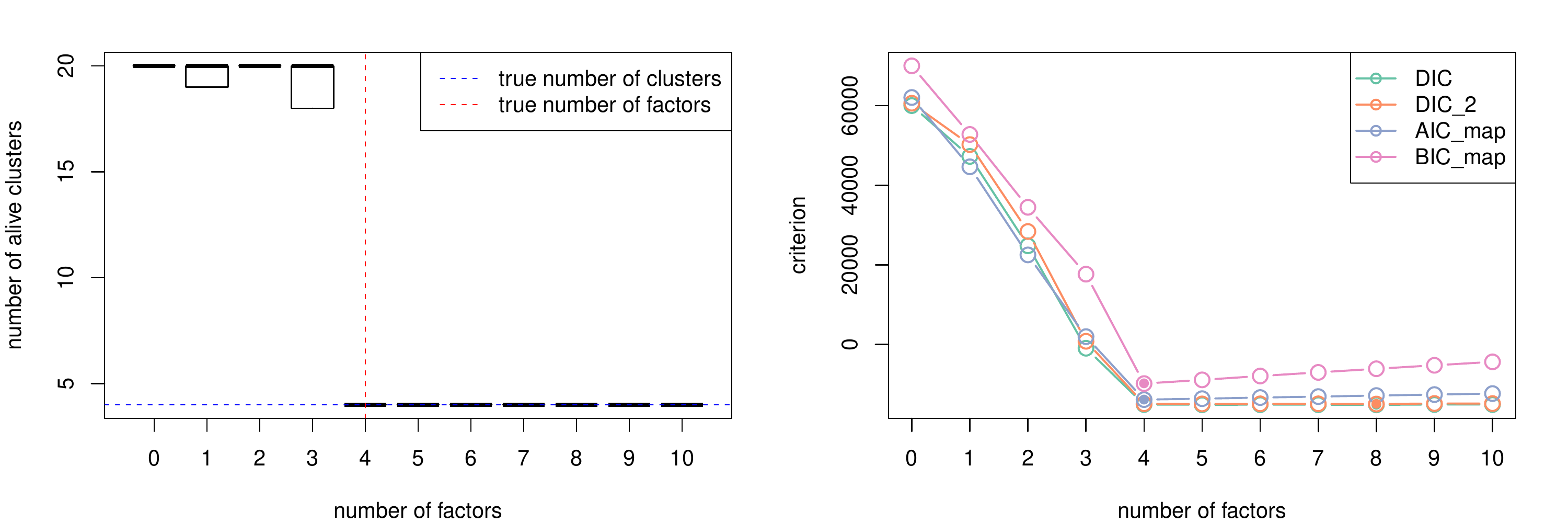}\\
\centering\includegraphics[scale=0.4]{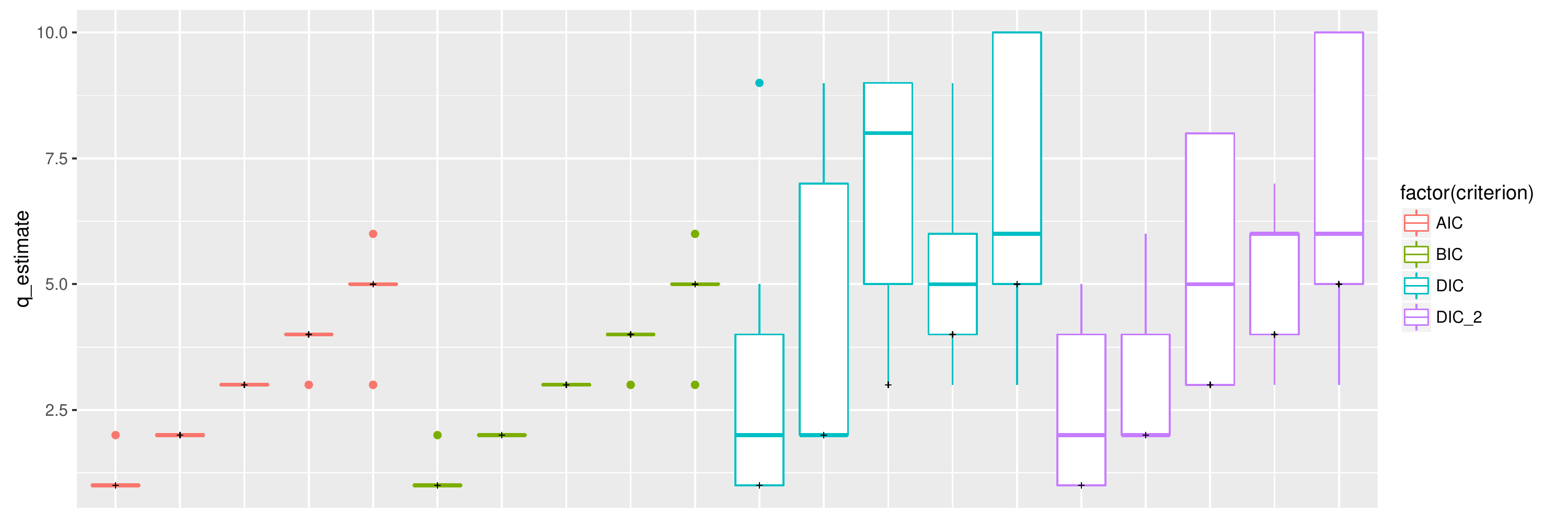}
\caption{Up left: Posterior distribution of the number of alive clusters considering various numbers of factors on a single dataset where the true number of clusters is $K = 4$ and the true number of factors is equal to $q = 4$. Up right: Estimated information criteria for each value of $q$, where a solid dot corresponds to the selected model. Down: Benchmarking the estimation of the number of factors on synthetic data using various model selection criteria. We considered that the number of factors ($q$) varies in the set $q\in\{1,\ldots,5\}$ and for each case 15 different datasets were simulated with number of clusters varying in the set $\{1,\ldots,10\}$. The cross denotes the real value of $q$.
}
\label{fig:k_versus_q}
\end{figure}

A typical output of the proposed method for a single dataset is shown in Figure \ref{fig:k_versus_q}. Note that we have also considered the case $q = 0$, which is explicitly taken into account into our method. Observe that when the number of factors is less than the true one ($q = 4$) the posterior distribution of the number of alive components supports  larger values than the true number of clusters ($K = 4$). As soon as the true number of factors is reached, then the posterior distribution of the number of alive clusters remains concentrated at the true value.

For each dataset, the number of factors was estimated by running our algorithm considering that $q\in\{0,1,\ldots,10\}$ and selecting the best model according to $\mbox{AIC}$, $\mbox{BIC}$, $\mbox{DIC}$ and $\mbox{DIC}_2$. As shown in Fig \ref{fig:k_versus_q} (down), both $\mbox{AIC}$ and $\mbox{BIC}$ are able to correctly estimate the number of factors. On the other hand, $\mbox{DIC}$ and $\mbox{DIC}_2$ tend to overfit, as often concluded in the literature (see e.g. \cite{spiegelhalter2014deviance}). However, even in the case that the Deviance Information Criterion is used for choosing the number of factors, the corresponding estimate of the underlying clusters (as well as the number of ``alive'' clusters) is  the same as the one returned by BIC or AIC. This behaviour is also illustrated in Figure \ref{fig:k_versus_q} (up) for a single dataset. Consequently, given that the MCMC sampler has converged, using the DIC or $\mbox{DIC}_2$ does not necessarily means that the inferred clustering is ``wrong''.

\bibliography{papastamoulis_revision2.bib}   

\end{document}